\documentstyle[preprint,tighten,aps,prd,eqsecnum]{revtex}
\newcommand{\clh}{{\mathcal{H}}}
\newcommand{\clp}{{\mathcal{P}}}
\newcommand{\cle}{{\mathcal{E}}}
\newcommand{\clb}{{\mathcal{B}}}

\newcommand{\cll}{{\mathcal{L}}}

\newcommand{\cly}{{\mathcal{Y}}}
\newcommand{\clz}{{\mathcal{Z}}}
\newcommand{\clq}{{\mathcal{Q}}}
\newcommand{\sigt}{\sigma_{\mbox{\scriptsize T}}}
\newcommand{\nelec}{n_{\text{e}}}
\newcommand{\TT}{\text{TT}}
\newcommand{\curl}{\,\text{curl}\,}
\newcommand{\uD}{\text{D}}
\newcommand{\ud}{\text{d}}
\begin{document}
\draft
\title{Microwave background polarization in cosmological models}
\author{Anthony Challinor\thanks{E-mail: A.D.Challinor@mrao.cam.ac.uk}}
\address{Astrophysics Group, Cavendish Laboratory, Madingley Road,
Cambridge CB3 0HE, UK.}
\date{\today}
\maketitle
\begin{abstract}
We introduce a new multipole formalism for polarized radiative transfer in
general spacetime geometries. The polarization tensor is expanded in
terms of coordinate-independent, projected symmetric trace-free (PSTF)
tensor-valued multipoles. The PSTF representation allows us to discuss
easily the observer dependence of the multipoles of the polarization, and to
formulate the exact dynamics of the radiation in convenient 1+3 covariant
form. For the case of an almost-Friedmann-Robertson-Walker (FRW) cosmological
model we recast the Boltzmann equation for the polarization in to a
hierarchy of multipole equations. This allows us to give a rigorous
treatment of the generation and propagation of the polarization of
the cosmic microwave background in almost-FRW models (with open, closed or
flat geometries) without recourse to any harmonic decomposition of the
perturbations. We also show how expanding the
intensity and polarization multipoles in derivatives of harmonic functions
gives a streamlined derivation of the mode-expanded multipole hierarchies.
Integral solutions to these hierarchies are provided, and the relation of
our formalism to others in the literature is discussed.
\end{abstract}

\section{Introduction}

With the growing body of data relating to the anisotropies of the cosmic
microwave background radiation (CMB), and the potential impact on cosmology of
conclusions drawn from the analysis of CMB data, there is a strong case
for developing a flexible, but physically transparent, formalism for
describing the propagation of the CMB in general cosmological models.

In a series of papers~\cite{maartens95,stoeger95b,dunsby96b,LC-sw,LC-scalcmb,%
gebbie98a,maartens99,chall99a,gebbie99b,chall99b}, a 1+3 covariant and
gauge-invariant formalism has been developed with a view to giving a
model-independent framework in which to study the physics of the CMB.
The approach is based on the projected symmetric trace-free (PSTF)
representation of relativistic kinetic theory due to Ellis, Treciokas, and
Matravers~\cite{ellis83}, and Thorne~\cite{thorne81}, and builds on the
covariant and gauge-invariant approach to perturbations in cosmology
(e.g. Ref.~\cite{ellis98}).
Some of the benefits of the covariant approach, which were emphasised in
Refs.~\cite{maartens99,gebbie99b}, include: (i) clarity in the definition
of the variables employed; (ii) covariant and gauge-invariant
perturbation theory around a variety of background models; (iii) provision
of a sound basis for studying non-linear effects; (iv) freedom to employ
any coordinate system or tetrad. In its current form, the covariant formalism
does not allow for the inclusion of polarization. It is this omission that we
address here, by introducing a new multipole formalism which allows one to
analyse CMB polarization in arbitrary cosmological models in the 1+3 covariant
approach.

We set up the multipole decomposition of the polarization and the exact
equations of radiative transfer (the Boltzmann equation) in convenient 1+3
covariant form. The polarization multipoles are covariantly-defined PSTF
tensors which leads to significant simplifications in the algebraic structure
of the scattering terms in the Boltzmann equation, and allows us to discuss
easily how the multipoles transform under changes of reference frame.
With the Boltzmann equation in 1+3 covariant form, the physical processes
responsible for the production and evolution of anisotropies and polarization
are particularly transparent. The PSTF multipole decomposition of the
polarization suggests recasting the Boltzmann equation as a multipole
hierarchy, in a similar manner to the treatment of the intensity in
Refs.~\cite{ellis83,thorne81}. This turns out to be quite involved, and we
shall only present the results for the special case
of an almost-FRW model here. (The exact, non-linear multipole equations will
be given in a subsequent paper.) These multipole equations allow us
to give a rigorous treatment of the generation and propagation of polarization
in the CMB in almost-FRW models for all spatial geometries.

Although the calculation of CMB anisotropies in an almost-FRW universe involves
only linear perturbation theory, the complexity of the subject is increased
when one allows for non-flat spatial geometries, vector and tensor
perturbations, and the inclusion of polarization~\cite{hu98}. To our
knowledge, the only published formalism for handling the most general
almost-FRW models is the total angular momentum method of
Hu et al.~\cite{hu98}. In their approach, considerable simplifications
result from the introduction of a normal mode expansion for the radiation
where the local angular and spatial distribution is explicit. In the 1+3
covariant approach the local angular distribution of the radiation
is analysed in terms of the PSTF tensor-valued multipoles, which can be done
without recourse to a decomposition of the spatial distribution in harmonic
functions. The multipole equations can then be used to analyse the
evolution of anisotropies and polarization for a quite general perturbation.
If required, the spatial dependencies can be handled by expanding the radiation
multipoles in derivatives of harmonic functions, which gives a streamlined
derivation of the mode-expanded multipole hierarchies.
Combining the angular and spatial expansions we find a normal mode
representation of the radiation which is equivalent to that of
Wilson~\cite{wilson83} for scalar temperature perturbations, while for vector
and tensor modes we obtain the obvious generalisation of Wilson's method.

The paper is arranged as follows. In Sec.~\ref{sec:multi} we introduce the PSTF
multipole decomposition of a polarized radiation field, relative to some
timelike velocity field $u^a$, obtaining the 1+3 covariant forms of the
electric and magnetic parts of the linear polarization tensor.
We also give an exact discussion of the transformation properties of the
polarization tensor under changes of the velocity field $u^a$, and give
non-linear expressions to first-order in relative velocities for the
transformations of the polarization multipoles.
In Sec.~\ref{sec:boltz} we discuss the Boltzmann equation for polarized
radiative transport. We give new, exact expressions for the scattering term in
the Thomson limit and the integral solution for the linear polarization,
valid in general cosmological models. Specialising to almost-FRW models, we
present the multipole
form of the Boltzmann equation which shows explicitly that the evolution of the
electric and magnetic parts of the linear polarization tensor are coupled
through curl terms, much like the coupling of electric and magnetic fields in
Maxwell's equations. In Sec.~\ref{sec:scalar} we decompose the multipole
equations into harmonic modes for scalar perturbations and provide the integral
solution which is central to the line of sight algorithm for efficient
numerical evaluation of the CMB power spectrum~\cite{seljak96}.
The analysis is repeated for tensor
perturbations in Sec.~\ref{sec:tens}, where a very direct derivation of the
integral solutions is also given. In Sec.~\ref{sec:disc} we
discuss the relation of our approach to others in the literature. Finally we
close with our conclusions in Sec.~\ref{sec:conc}. An appendix summarises
the PSTF representation of the scalar and tensor harmonics.

We employ a $(+---)$ signature for the spacetime metric $g_{ab}$. Early lower
case Roman indices $a$, $b$ etc. refer to a general basis, with $i$ and $j$
referring to the 1, 2 components in an orthonormal tetrad,
$(e_i)^a$, $i=1,2$. Round brackets denote symmetrisation on the
enclosed indices, square brackets antisymmetrisation, and angle brackets the
PSTF part: $S_{\langle ab \rangle} \equiv (h_{(a}^c h_{b)}^d - \case{1}{3}
h_{ab} h^{cd})S_{cd}$, where the projection operator $h^{ab}
\equiv g^{ab}-u^a u^b$ with $u^a$ the fundamental velocity of the 1+3
covariant approach. The index notation $A_l$ denotes the index string
$a_1\dots a_l$, and the notation $e_{A_l}$ denotes the tensor product
$e_{a_1}\dots e_{a_l}$. We use units with $c=G=1$ throughout.

\section{Multipole decomposition of the radiation field}
\label{sec:multi}

We describe observations from the viewpoint of an observer comoving with
the fundamental velocity field $u^a$. In a general cosmological model there
is some freedom in the choice of $u^a$, although we must ensure that
$u^a$ is defined in a physical manner (such as by the timelike eigenvector
of the matter stress-energy tensor, or the 4-velocity of some particle species)
so that in the FRW limit, $u^a$ correctly reduces to the fundamental velocity
of the FRW model. This restriction on $u^a$ is necessary to ensure
gauge-invariance of the 1+3 covariant perturbation theory. A photon with
4-momentum $p^a$ has an energy $E$ and propagation direction $e^a$ relative to
$u^a$, where
\begin{equation}
p^a = E(u^a + e^a).
\label{eq:1}
\end{equation}
For a given propagation direction $e^a$, the observer can introduce a
pair of orthogonal polarization vectors $(e_1)^a$ and $(e_2)^a$
which are perpendicular to $u^a$ and $e^a$:
\begin{equation}
\clh_b^a (e_1)^b = (e_1)^a,
\label{eq:2}
\end{equation}
with a similar result for $(e_2)^a$. Here $\clh_{ab}$ is the screen projection
tensor that projects perpendicular to both $u^a$ and $e^a$:
\begin{equation}
\clh_{ab} = h_{ab} + e_a e_b.
\label{eq:3}
\end{equation}
We shall refer to tensors like $(e_1)^a$, which are perpendicular to $e^a$
and $u^a$, as being transverse.
The set of vectors $\{ u^a, (e_1)^a, (e_2)^a, e^a \}$ form a right-handed
orthonormal tetrad at the observation point. Using the polarization basis
vectors, the observer can decompose an arbitrary radiation field into
Stokes parameters (e.g. Ref.~\cite{chand_rad}) $I(E,e^a)$, $Q(E,e^a)$,
$U(E, e^a)$ and $V(E,e^a)$ along the direction $e^a$ at photon energy (or
equivalently frequency) $E$. The transformation laws of the Stokes
parameters under rotations of $(e_1)^a$ and $(e_2)^a$ lead one to introduce
a second-rank transverse polarization tensor $P_{ab}(E,e^c)$. The
only non-vanishing tetrad components of $P_{ab}(E,e^c)$ are
\begin{equation}
P_{ab}(e_i)^a (e_j)^b = \frac{1}{2} \left( \begin{array}{cc}
I + Q & U + V \\
U - V & I- Q \end{array} \right),
\label{eq:4}
\end{equation}
for $i$ and $j=1,2$, and we have left the arguments $E$ and $e^a$
implicit. Introducing the projected alternating tensor $\epsilon_{abc}
\equiv \eta_{abcd} u^d$, where $\eta_{abcd}$ is the spacetime
alternating tensor, we can write $P_{ab}(E,e^c)$ in the covariant irreducible
form
\begin{equation}
P_{ab}(E,e^d) = -\frac{1}{2}I(E,e^d) \clh_{ab} + \clp_{ab}(E,e^d)
+\frac{1}{2}V(E,e^d) \epsilon_{abc}e^c,
\label{eq:5}
\end{equation}
which defines the linear polarization tensor $\clp_{ab}(E,e^c)$, where
\begin{equation}
\clp_{ab}(e_i)^a (e_j)^b = \frac{1}{2} \left( \begin{array}{cc}
Q & U \\
U & - Q \end{array} \right),
\label{eq:6}
\end{equation}
which is transverse and trace-free. For a general second-rank tensor
$S_{ab}$, we follow Thorne~\cite{thorne80} by denoting the transverse
trace-free (TT) part by $[S_{ab}]^{\TT}$, so that
\begin{equation}
[S_{ab}]^{\TT} = \clh_a^c \clh_b^d S_{cd} - \frac{1}{2} \clh_{ab}
\clh^{cd} S_{cd}.
\label{eq:7}
\end{equation}
The magnitude (squared) of the linear polarization is $Q^2+U^2$ which
can be written in the manifestly basis-independent form $2\clp_{ab}\clp^{ab}$.

Since $I(E,e^c)$ and $V(E,e^c)$ are scalar functions on the
sphere at a point in spacetime, their local angular dependence can be handled
by an expansion in PSTF tensor-valued multipoles~\cite{ellis83,thorne81}:
\begin{eqnarray}
I(E,e^c) & = & \sum_{l=0}^\infty I_{A_l}(E) e^{A_l} , \label{eq:8}\\
V(E,e^c) & = & \sum_{l=0}^\infty V_{A_l}(E) e^{A_l} . \label{eq:9}
\end{eqnarray}
The expansion in PSTF multipoles is the coordinate-free version
of the familiar spherical harmonic expansion. Eqs.~(\ref{eq:8})
and (\ref{eq:9}) can be inverted using the orthogonality of the
$e^{\langle A_l \rangle}$, for example,
\begin{equation}
I_{A_l}(E) = \Delta_l{}^{-1} \int \ud\Omega \,I(E,e^c) e_{\langle A_l \rangle}
\qquad \text{where} \qquad \Delta_l \equiv \frac{4\pi(-2)^l(l!)^2}{(2l+1)!}.
\label{eq:10}
\end{equation}

For the TT tensor $\clp_{ab}(E,e^c)$, we use the TT elements of the PSTF
representation of the pure-spin tensor spherical harmonics
(see Ref.~\cite{thorne80}
for a thorough review of the various representations of the tensor spherical
harmonics), so that
\begin{equation}
\clp_{ab}(E,e^c) = \sum_{l=2}^\infty[\cle_{ab C_{l-2}}(E)
e^{C_{l-2}}]^{\TT} + \sum_{l=2}^\infty
[e_{d_1}\epsilon^{d_1 d_2}{}_{(a}\clb_{b)d_2 C_{l-2}}(E)
e^{C_{l-2}}]^{\TT}.
\label{eq:11}
\end{equation}
The first summation in Eq.~(\ref{eq:11}) defines the electric part of the
linear polarization, while the second summation defines the magnetic part.
(The electric and magnetic parts of the polarization were denoted by
G(radient) and C(url) in Ref.~\cite{kamion97}.) The $l$-th term in the
multipole expansion of the electric part has parity $(-1)^l$, while
the $l$-th term in the magnetic part has parity $(-1)^{l+1}$.
Eq.~(\ref{eq:11}) can be inverted to give
\begin{eqnarray}
\cle_{A_l}(E) & = & M_l{}^2 \Delta_l{}^{-1}
\int \ud\Omega\, e_{\langle A_{l-2}} \clp_{a_{l-1} a_l \rangle} (E,e^c) ,
\label{eq:12} \\
- \clb_{A_l}(E) & = & M_l{}^2 \Delta_l{}^{-1}
\int \ud\Omega\, e_b \epsilon^{bd}{}_{\langle a_l} e_{A_{l-2}}
\clp_{a_{l-1}\rangle d} (E,e^c), \label{eq:13}
\end{eqnarray}
where $M_l \equiv \sqrt{2l(l-1)/[(l+1)(l+2)]}$.
The multipole expansion in Eq.~(\ref{eq:11}) is the coordinate-free version
of the tensor spherical harmonic expansion introduced to the analysis of
CMB polarization in Ref.~\cite{kamion97}, and the expansion in
the Newman-Penrose spin-weight 2 spherical harmonics employed in
Ref.~\cite{seljak97}.

\subsection{Power spectra}
\label{subsec:power}

To define the bolometric power spectrum it is convenient to define
energy-integrated multipoles, for example~\footnote{$I_{A_l}$ was
denoted $\Delta_l \Pi_{A_l}$ in Ref.~\cite{maartens99}, and
$J^{(l)}_{A_l}$ in Ref.~\cite{LC-scalcmb}.}
\begin{equation}
I_{A_l} = \Delta_l \int_0^\infty \ud E\, I_{A_l}(E),
\label{eq:14}
\end{equation}
with equivalent definitions for $\cle_{A_l}$, $\clb_{A_l}$ and
$V_{A_l}$.
The factor $\Delta_l$ is included in Eq.~(\ref{eq:14})
so that the three lowest intensity multipoles give the radiation energy
density, energy flux and anisotropic stress respectively:
\begin{equation}
I=\rho^{(\gamma)}, \qquad I_a = q_a^{(\gamma)}, \qquad I_{ab} =
\pi_{ab}^{(\gamma)}.
\label{eq:15}
\end{equation}
In the almost-FRW case, with an ensemble that is statistically isotropic,
the power spectrum of the (bolometric) temperature anisotropies, $C_l^{II}$,
can be defined in terms of the covariance of the intensity
multipoles~\cite{gebbie98a}:
\begin{equation}
\left(\frac{\pi}{I}\right)^2 \langle I_{A_l} I^{B_{l'}}\rangle
= \Delta_l C_l^{II}\delta_l^{l'} h_{\langle A_l \rangle}^{\langle B_l \rangle},
\label{eq:16}
\end{equation}
where $h_{\langle A_l \rangle}^{\langle B_l \rangle}\equiv
h_{\langle a_1}^{\langle b_1}\dots h_{a_l \rangle}^{b_l \rangle}$.
For small anisotropies, the gauge-invariant fractional temperature difference
from the all-sky mean is related to the $I_{A_l}$ by
\begin{equation}
\delta_T(e^c) = \frac{\pi}{I} \sum_{l=1}^\infty \Delta_l{}^{-1}
I_{A_l} e^{A_l},
\label{eq:17}
\end{equation}
so that the temperature correlation function evaluates to
\begin{equation}
\langle \delta_T(e^c) \delta_T(e^{\prime c}) \rangle =
\sum_{l=1}^\infty \frac{(2l+1)}{4\pi} C_l^{II} P_l(X),
\label{eq:18}
\end{equation}
where $X\equiv -e^a e_a^\prime$, and $P_l(X)$ is a Legendre polynomial.
In deriving Eq.~(\ref{eq:18}) we used the result $e^{\langle A_l \rangle}
e'_{\langle A_l \rangle} = (2l+1)\Delta_l P_l(X)/(4\pi)$.

The power spectra for the polarization are defined by analogy with
Eq.~(\ref{eq:16}). We choose our conventions for the power spectra to conform
with Ref.~\cite{seljak97}, so that it is necessary to include an additional
factor of $M_l/\sqrt{2}$ for each factor of the polarization.
For example, for the electric polarization
\begin{equation}
\left(\frac{\pi}{I}\right)^2 \langle \cle_{A_l} \cle^{B_{l'}}\rangle
= \frac{l(l-1)}{(l+1)(l+2)}\Delta_l C_l^{\cle\cle} \delta_l^{l'}
h_{\langle A_l \rangle}^{\langle B_l \rangle}.
\label{eq:19}
\end{equation}
Unfortunately, the definitions of the polarization power spectra given in
Ref.~\cite{seljak97} differ from those in Ref.~\cite{kamion97} by factors of
$\sqrt{2}$; see Ref.~\cite{kamion97c} and Sec.~\ref{sec:disc} for details.
For a parity symmetric ensemble the magnetic polarization is uncorrelated with
the electric polarization and the temperature anisotropies. In this case,
we find that the generalisation of Eq.~(\ref{eq:18}) to linear polarization is
\begin{equation}
2\left(\frac{\pi}{I}\right)^2 \langle \clp_{ab}(e^c) \clp^{ab}(e^{\prime c})
\rangle = \sum_{l=2}^\infty \frac{(2l+1)}{4\pi} [C_l^{\cle \cle} P_{l-2}(X)
+C_l^{\clb\clb} P_{l-1}(X) ],
\label{eq:20}
\end{equation}
where $\clp_{ab}(e^c)=\int_0^\infty \ud E\, \clp_{ab}(E,e^c)$.
When the directions $e^c$ and $e^{\prime c}$ coincide, Eq.~(\ref{eq:20})
reduces to the ensemble average of the square of the degree of linear
polarization (expressed as a dimensionless bolometric temperature).
Note that the electric
polarization quadrupole gives an isotropic contribution to the
linear polarization correlation function. Integrating Eq.~(\ref{eq:20})
over all directions $e^{\prime c}$, with $e^c$ fixed, the right-hand
side reduces to $5C_2^{\cle\cle}$. The presence of $P_{l-2}(X)$ and
$P_{l-1}(X)$  in Eq.~(\ref{eq:20}) reflects the opposite parities of the
electric and magnetic contributions to $\clp_{ab}(e^c)$ for a given multipole
$l$.

\subsection{Transformation laws under changes of frame}
\label{subsec:trans}

The polarization tensor $P_{ab}(E,e^c)$ has been expressed in 1+3
covariant form, so it is not invariant under changes of frame. If we consider
a new velocity field $\tilde{u}^a=\gamma(u^a+v^a)$, where $v^a$ is the
projected relative velocity in the $u^a$ frame and $\gamma$ is the associated
Lorentz factor, for a given photon with
4-momentum $p^a$ the energy and propagation directions in the
$\tilde{u}^a$ frame are given by the Doppler and aberration formulae:
\begin{eqnarray}
\tilde{E} & = & \gamma E (1+e^a v_a), \label{eq:21} \\
\tilde{e}^a & = & [\gamma(1+e^b v_b)]^{-1}(u^a + e^a)-\gamma(u^a+v^a).
\label{eq:22}
\end{eqnarray}
Note that $\tilde{e}^a$ is a projected vector relative to $\tilde{u}^a$.
The screen projection tensor for a given null direction transforms to
\begin{equation}
\tilde{\clh}_{ab} = \clh_{ab} - \frac{2\gamma}{\tilde{E}}
p_{(a}\clh_{b)c}v^c + \frac{\gamma^2}{\tilde{E}^2}p_a p_b \clh_{cd}
v^c v^d,
\label{eq:23}
\end{equation}
while the polarization tensor transforms according to
\begin{equation}
\tilde{E}^{-3}\tilde{P}_{ab}(\tilde{E},\tilde{e}^c)=E^{-3}\tilde{\clh}_a^{d_1}
\tilde{\clh}_b^{d_2} P_{d_1 d_2}(E,e^c).
\label{eq:24}
\end{equation}
Under this transformation law the intensity, circular polarization and
linear polarization do not mix. The irreducible components of Eq.~(\ref{eq:24})
give the frame invariance of $I(E,e^c)/E^3$ and $V(E,e^c)/E^3$, and show that
the transformation law for $\clp_{ab}(E,e^c)$ follows that for
$P_{ab}(e^c)$. The degree of linear polarization
$[2\clp_{ab}(E,e^c)\clp^{ab}(E,e^c)]^{1/2}/I(E,e^c)$ is invariant under
changes of frame. The transformation law for $P_{ab}(E,e^c)$ ensures that
the tetrad components of $P_{ab}(E,e^c)/E^3$ are invariant if the
polarization basis vectors are transformed as
\begin{equation}
(\tilde{e}_i)^a = \tilde{\clh}^a_b (e_i)^b,
\label{eq:25}
\end{equation}
for $i=1,2$.

Under changes of frame, multipoles with different $l$ mix because of
Doppler beaming effects. To first-order in the relative velocity
$v^a$, the energy-integrated multipoles transform as
\begin{equation}
\tilde{I}_{A_l} = \tilde{h}^{\langle B_l \rangle}_{\langle A_l \rangle}
I_{B_l} - (l-2) v^b I_{b A_l} - \frac{l(l+3)}{(2l+1)} v_{\langle
a_l}I_{A_{l-1}\rangle},
\label{eq:26}
\end{equation}
with an equivalent result for $V_{A_l}$. Although the linear polarization
transforms irreducibly, the electric and magnetic parts do mix. To
first-order in $v^a$, we find
\begin{eqnarray}
\tilde{\cle}_{A_l} &=& \tilde{h}^{\langle B_l \rangle}_{\langle A_l \rangle}
\cle_{B_l} - \frac{l(l+3)}{(2l+1)} v_{\langle a_l}\cle_{A_{l-1}\rangle}
- \frac{(l-2)(l-1)(l+3)}{(l+1)^2} v^b \cle_{b A_l} \nonumber \\
&&\mbox{}-\frac{6}{(l+1)}
v_b\epsilon^{bc}{}_{\langle a_l}\clb_{A_{l-1}\rangle c},
\label{eq:27}\\
\tilde{\clb}_{A_l} &=& \tilde{h}^{\langle B_l \rangle}_{\langle A_l \rangle}
\clb_{B_l} - \frac{l(l+3)}{(2l+1)} v_{\langle a_l}\clb_{A_{l-1}\rangle}
- \frac{(l-2)(l-1)(l+3)}{(l+1)^2} v^b \clb_{b A_l} \nonumber \\
&&\mbox{} +\frac{6}{(l+1)}
v_b\epsilon^{bc}{}_{\langle a_l}\cle_{A_{l-1}\rangle c}.
\label{eq:28}
\end{eqnarray}
In an almost-FRW model the polarization is a first-order quantity, as
are physically defined relative velocities. It follows that the electric and
magnetic multipoles are frame-invariant in linear theory.

\section{Boltzmann equation}
\label{sec:boltz}

The dynamics of the radiation field is described by the (exact)
collisional Boltzmann equation
\begin{equation}
\cll [E^{-3}P_{ab}(E,e^c)]=K_{ab}(E,e^c),
\label{eq:29}
\end{equation}
where the Liouville operator $\cll$ acts on transverse tensors $A_{ab}=[A_{ab}
]^{\TT}$ along the photon path $x^a(\lambda)$,
$p^a(\lambda)$ in phase space, with $p^a = \ud x^a/\ud\lambda$, according to
\begin{equation}
\cll[A_{ab}(E,e^c)] \equiv \clh_a^{d_1} \clh_b^{d_2} p^e \nabla_e A_{d_1 d_2}
(E,e^c),
\label{eq:30}
\end{equation}
where $\nabla_a$ is the spacetime covariant derivative. The transverse
scattering tensor $K_{ab}(E,e^c)$ describes interactions of the radiation with
matter. The Liouville operator $\cll$ preserves the
irreducible decomposition of the polarization tensor [Eq.~(\ref{eq:5})],
so that
\begin{eqnarray}
\cll[E^{-3}P_{ab}(E,e^c)] &=& -\frac{1}{2} \frac{\ud}{\ud\lambda}
[E^{-3}I(E,e^c)]\clh_{ab} + \cll[E^{-3}\clp_{ab}(E,e^c)] \nonumber \\
&&\mbox{}+\frac{1}{2} \frac{\ud}{\ud\lambda}[E^{-3}V(E,e^c)]\epsilon_{abd}e^d.
\label{eq:31}
\end{eqnarray}
In the absence of scattering, $K_{ab}(E,e^c)=0$, the tetrad components of
$P_{ab}(E,e^c)/E^3$ are constant along the photon path provided that the
polarization basis vectors are transported along the null geodesic according to
\begin{equation}
\clh_{ab} p^c \nabla_c (e_i)^b =0,
\label{eq:32}
\end{equation}
and the constraint $(e_i)^a = \clh^a_b (e^i)^b$.

It follows from Eqs.~(\ref{eq:24}) and (\ref{eq:30}) that under changes of
frame $u^a\mapsto \tilde{u}^a$, the action of the Liouville
operator transforms as
\begin{equation}
\tilde{\cll}[\tilde{E}^{-3}\tilde{P}_{ab}(\tilde{E},\tilde{e}^c)]
=\tilde{\clh}_a^{d_1} \tilde{\clh}_b^{d_2} \cll[E^{-3}P_{d_1 d_2}(E,e^c).
\label{eq:33}
\end{equation}
With this result, we deduce from Eq.~(\ref{eq:29}) that the scattering
tensor $K_{ab}(E,e^c)$ must transform as
\begin{equation}
\tilde{K}_{ab}(\tilde{E},\tilde{e}^c)=\tilde{\clh}_a^{d_1} \tilde{\clh}_b^{d_2}
K_{d_1 d_2}(E,e^c).
\label{eq:34}
\end{equation}
This result is useful since the scattering tensor is often simplest to
evaluate in some preferred frame, picked out by the physics of the scattering
process. The scattering tensor in a general frame can then be computed
using Eq.~(\ref{eq:34}).

Over the epochs relevant to the formation of anisotropies and polarization
in the CMB the dominant coupling between the radiation and the matter is
Compton scattering. To an excellent approximation we can ignore the effects
of Pauli blocking, induced scattering, and electron recoil in the rest frame
of the scattering electron, so that the scattering may be approximated by
classical Thomson scattering in the electron rest frame. Taking the
electron 4-velocity to be $\tilde{u}^a$, and the (proper) number density of
free electrons to be $\tilde{n}_{{\text{e}}}$, the exact scattering tensor
in the $\tilde{u}^a$ frame in the classical Thomson limit is
\begin{eqnarray}
\tilde{E}^2 \tilde{K}_{ab}(\tilde{E},\tilde{e}^c) &=& \tilde{n}_{{\text{e}}}
\sigt \biggl\{-\frac{1}{2} \tilde{\clh}_{ab}
\left[-\tilde{I}(\tilde{E},\tilde{e}^c) + \tilde{I}(\tilde{E})
+ {\frac{1}{10}} \tilde{I}_{d_1 d_2}(\tilde{E})\tilde{e}^{d_1} \tilde{e}^{d_2}
+ {\frac{3}{5}} \tilde{\cle}_{d_1 d_2}(\tilde{E}) \tilde{e}^{d_1}
\tilde{e}^{d_2}\right] \nonumber \\
&&\mbox{}+ \left[- \tilde{\clp}_{ab}(\tilde{E},\tilde{e}^c)
+ {\frac{1}{10}} [\tilde{I}_{ab}(\tilde{E})]^{\TT}
+ {\frac{3}{5}} [\tilde{\cle}_{ab}(\tilde{E})]^{\TT} \right] \nonumber \\
&&\mbox{}+ {\frac{1}{2}} \tilde{\epsilon}_{abd_1} \tilde{e}^{d_1}
\left[- \tilde{V}(\tilde{E},\tilde{e}^c) + {\frac{1}{2}} \tilde{V}_{d_2}
(\tilde{E}) \tilde{e}^{d_2}\right]\biggr\},
\label{eq:35}
\end{eqnarray}
where $\sigt$ is the Thomson cross section.
This expression for the scattering tensor follows from inserting the
multipole decomposition of the polarization tensor into the Kernel for
Thomson in-scattering (e.g. Ref.~\cite{chand_rad}), and integrating over
scattering directions.

We have written Eq.~(\ref{eq:35}) in irreducible form.
The first set of terms in square brackets affects the evolution of the
intensity in the electron rest frame, the second set of terms affect the
linear polarization, and the third set affect the circular polarization.
Concentrating first on the terms that affect the intensity, the
lack of a monopole component is due to there being no energy transfer from
Thomson scattering in the rest frame of the electron, while the presence of
the intensity and electric polarization quadrupoles is due to the
anisotropy and polarization dependence of Thomson scattering.
Turning to the terms in $\tilde{K}_{ab}(\tilde{E},\tilde{e}^c)$ which
couple to the evolution of the linear polarization, the presence of
the quadrupoles of the intensity and the electric polarization arises
from Thomson in-scattering. These terms directly affect the
evolution of the electric quadrupole alone in the electron rest frame, but this
is not true in a general frame. Finally, we see from the terms in the final
square bracket in Eq.~(\ref{eq:35}) that the evolution of the circular
polarization decouples from the intensity and the linear polarization.
Since the circular polarization transforms irreducibly under changes of frame,
circular polarization is not sourced by Thomson scattering for
arbitrary frame choices.

\subsection{Integral solution for {$\clp_{ab}(e^c)$}}

The transformation of $\tilde{K}_{ab}(\tilde{E},\tilde{e}^c)$ to a general
frame gives a non-local expression in the energy $E$, since there is
energy transfer in Thomson scattering from a moving electron.
(See Ref.~\cite{CFL99} for applications of this transformation in
cluster physics.) It is therefore
convenient to integrate over energy, in which case we find the following
exact expressions for the evolution along the line of sight:
\begin{eqnarray}
\int_0^\infty \ud E\, E^2 \frac{\ud}{\ud \lambda}[E^{-3} I(E,e^c)] &=&
- \tilde{n}_{\text{e}} \sigt \gamma (1+ v_d e^d) I(e^c) \nonumber \\
&&\mbox{}+ {\frac{1}{4\pi}} \tilde{n}_{\text{e}} \sigt
[\gamma(1+v_d e^d)]^{-3}(\tilde{I}+\tilde{\zeta}_{ab}\tilde{e}^a \tilde{e}^b),
\label{eq:36} \\
\int_0^\infty \ud E\, E^2 \cll[E^{-3} \clp_{ab}(E,e^c)] &=& -
\tilde{n}_{\text{e}} \sigt \gamma (1+ v_d e^d) \clp_{ab}(e^c) \nonumber \\
&&\mbox{}+ {\frac{1}{4\pi}} \tilde{n}_{\text{e}} \sigt [\gamma(1+v_d e^d)]^{-3}
[\tilde{\clh}^{c_1}_a \tilde{\clh}^{c_2}_b \tilde{\zeta}_{c_1 c_2}]^{\TT},
\label{eq:37} \\
\int_0^\infty \ud E\, E^2 \frac{\ud}{\ud\lambda}[E^{-3} V(E,e^c)]
&=& -\tilde{n}_{\text{e}} \sigt \gamma (1+ v_d e^d) V(e^c) \nonumber \\
&&\mbox{} - {\frac{3}{8\pi}}\tilde{n}_{\text{e}} \sigt [\gamma(1+v_d e^d)]^{-3}
\tilde{V}_a \tilde{e}^a.
\label{eq:38}
\end{eqnarray}
The tildes denote that the quantity is evaluated in the rest frame of the
electrons, which has four-velocity $\tilde{u}^a = \gamma(u^a+v^a)$, and we
have defined e.g. $I(e^c)\equiv \int_0^\infty \ud E I(E,e^c)$.
The quantity $\zeta_{ab}$, where
\begin{equation}
\zeta_{ab} \equiv \frac{3}{4} I_{ab} + \frac{9}{2} \cle_{ab},
\label{eq:39}
\end{equation}
appears in Eqs.~(\ref{eq:36}) and (\ref{eq:37}), evaluated in the $\tilde{u}^a$
frame, because of the anisotropy and polarization dependence of Thomson
scattering.

Formal solutions to Eqs.~(\ref{eq:36}--\ref{eq:38}) can be obtained by
integrating along the null cone. For the linear polarization we must
work with the tetrad components of $\clp_{ab}(e^c)$, introduced in
Sec.~\ref{sec:multi}, since we can only meaningfully integrate scalar
equations. Making use of
\begin{eqnarray}
(e_i)^a (e_j)^b \int_0^\infty \ud E\, E^2 \cll[E^{-3}\clp_{ab}(E,e^c)]
&=& (e^a+u^a)\nabla_a [\clp_{ab}(e^c) (e_i)^a (e_j)^b] \nonumber \\
&&\mbox{} +(e_i)^a (e_j)^b \int_0^\infty \ud E\, E^2 \frac{\ud E}{\ud\lambda}
\frac{\partial}{\partial E} [E^{-3} \clp_{ab}(E,e^c)],
\label{eq:40}
\end{eqnarray}
and integrating the second term on the right by parts, we find the following
exact solution for the energy-integrated linear polarization tensor at some
point $R$ in a general cosmological model:
\begin{equation}
\clp_{ab}(e^c) (e_i)^a (e_j)^b |_R = \frac{1}{4\pi} \int^R \ud t\,
\tilde{n}_{\text{e}}\sigt e^{-\tilde{\tau}} (1+z)^{-4} [\gamma(1+v_d e^d)]^{-3}
[\tilde{\clh}_a^{c_1} \tilde{\clh}_b^{c_2} \tilde{\zeta}_{c_1 c_2}]^{\TT}
(e_i)^a (e_j)^b.
\label{eq:41}
\end{equation}
Here, the integral is along the photon geodesic and the measure is
$\ud t \equiv \ud x^a u_a$. The redshift back from $R$ along the line of
sight is $z$, and $\tilde{\tau}$ is the optical depth along the line of sight,
measured in the electron rest frame:
\begin{equation}
\tilde{\tau}|_A \equiv \int_A^R \ud \tilde{t} \, \tilde{n}_{\text{e}} \sigt,
\label{eq:42}
\end{equation}
where $\ud \tilde{t} \equiv \tilde{u}_a \ud x^a$, and $A$ is a point on
the null geodesic through $R$. The integral solution, Eq.~(\ref{eq:41}),
shows clearly how the intensity and electric quadrupoles generate
polarization on scattering, and the effect of subsequent
gravitationally-induced rotation of the basis vectors $(e_i)^a$ as the
radiation free streams to us. The need to calculate the intensity and electric
quadrupoles, the redshift, and the rotation of the basis vectors along the line
of sight suggests that in a general (non-linear) model it will be more
convenient to work directly with the polarization multipoles. However,
in Sec.~\ref{sec:tens} we show that the linearised version of
Eq.~(\ref{eq:41}) does provide a very direct route through to the integral
solutions for the mode-expanded multipoles in the almost-FRW case.

\subsection{Multipole decomposition of the {B}oltzmann equation}
\label{subsec:multibol}

The Boltzmann equation~(\ref{eq:29}) can be written in multipole form by
expressing $P_{ab}(E,e^c)$ as a multipole expansion using
Eqs.~(\ref{eq:8}), (\ref{eq:9}), and (\ref{eq:11}), and decomposing the
resulting equation into multipoles. This leads to four sets of multipole
hierarchies for $I_{A_l}(E)$, $\cle_{A_l}(E)$, $\clb_{A_l}(E)$,
and $V_{A_l}(E)$. In the absence of scattering the hierarchies for
$I_{A_l}(E)$ and $V_{A_l}(E)$ decouple, but if Thomson scattering is
included the hierarchy for $I_{A_l}(E)$ includes the multipoles of the
linear polarization as source terms. The exact multipole equations for
$I_{A_l}(E)$ in the absence of scattering were given in
Refs.~\cite{maartens99,ellis83,thorne81}; the same equations apply to the
multipoles $V_{A_l}$ of the circular polarization when there is no
scattering. (The leading non-linear
Thomson source terms in the hierarchy for the energy-integrated multipoles
$I_{A_l}$ were also given in Ref.~\cite{maartens99}, but these did
not include polarization effects.) The exact multipole equations
for $\cle_{A_l}(E)$ and $\clb_{A_l}(E)$ are significantly more
involved than those for the intensity or circular polarization. We defer
the derivation of the exact polarization multipole equations to a subsequent
paper. Here, we shall only consider the case of almost-FRW universes,
in which case the polarization and $l > 0$ intensity multipoles
are $O(\epsilon)$ in a smallness parameter, $\epsilon$, which characterises
the departure of the cosmological model from exact FRW symmetry. For the rest
of this paper we shall restrict attention to almost-FRW models, and work only
to first-order in $\epsilon$ (linear perturbation theory).

To first-order in $\epsilon$, we find for the electric
polarization\footnote{We adopt the convention that $\cle_{A_l}$ and
$\clb_{A_l}$ vanish for $l < 2$, and $I_{A_l}$ and $V_{A_l}$ vanish for $l<0$.}
\begin{eqnarray}
\dot{\cle}_{A_l}(E) &-& \frac{1}{3}\Theta E^4\frac{\partial}{\partial E}
[E^{-3}\cle_{A_l}(E)] + \uD_{\langle a_{l}}\cle_{A_{l-1}\rangle}
(E) - \frac{(l+3)(l-1)}{(l+1)(2l+3)} \uD^b \cle_{b A_l}(E)
\nonumber \\
&-& \frac{2}{(l+1)} \curl \clb_{A_l}(E) = -\nelec \sigt \cle_{A_l}(E)
+ \frac{1}{10}\nelec \sigt[I_{a_1 a_2}(E) + 6 \cle_{a_1 a_2}(E)]\delta_l^2,
\label{eq:43}
\end{eqnarray}
and for the magnetic polarization
\begin{eqnarray}
\dot{\clb}_{A_l}(E) &-& \frac{1}{3}\Theta E^4\frac{\partial}{\partial E}
[E^{-3}\clb_{A_l}(E)] + \uD_{\langle a_l}\clb_{A_{l-1}\rangle}
(E) - \frac{(l+3)(l-1)}{(l+1)(2l+3)} \uD^b \clb_{b A_l}(E)
\nonumber \\
&+& \frac{2}{(l+1)} \curl \cle_{A_l}(E) = -\nelec \sigt \clb_{A_l}(E).
\label{eq:44}
\end{eqnarray}
In these equations, an overdot denotes the action of $u^a\nabla_a$,
$\uD_a$ is the totally projected derivative:
\begin{equation}
\uD_a S_{b \dots c} \equiv h_a^d h_b^e \dots h_c^f \nabla_d S_{e \dots f},
\label{eq:45}
\end{equation}
for some arbitrary tensor $S_{b\dots c}$, and the volume expansion
$\Theta \equiv \nabla_a u^a$. We have replaced $\tilde{n}_{\text{e}}$ by the
electron number density in the $u^a$ frame, which is correct to the required
order. The electric and magnetic multipoles are
coupled through curl terms, where we have introduced the curl of a rank-$l$
PSTF tensor $S_{A_l}$,
\begin{equation}
\curl S_{A_l} \equiv \epsilon_{bc \langle a_l} \uD^b S_{A_{l-1}
\rangle}{}^c.
\label{eq:46}
\end{equation}
The coupling of electric and magnetic multipoles
through curl terms is reminiscent of Maxwell's equations, as is the fact that
only the electric multipoles have inhomogeneous source terms on the right.
It is this curl coupling and the structure of the source terms that lead
to the well-known result that
scalar perturbations do not generate magnetic polarization
(see Sec.~\ref{sec:scalar}). This is critical for the detection of a
gravitational wave component to the anisotropy~\cite{kamion98}.

It is often more convenient to work with the energy-integrated multipoles
$\cle_{A_l}$ and $\clb_{A_l}$, since it follows from Eqs.~(\ref{eq:43}) and
(\ref{eq:44}) that the polarization has the same energy spectrum as the
intensity anisotropies. Integrating over energies and then by parts, we find
\begin{eqnarray}
\dot{\cle}_{A_l} &+&\frac{4}{3}\Theta \cle_{A_l} + \frac{(l+3)(l-1)}{(l+1)^2}
\uD^b \cle_{b A_l} - \frac{l}{(2l+1)} \uD_{\langle a_l}\cle_{A_{l-1}\rangle}
\nonumber \\
&-& \frac{2}{(l+1)} \curl \clb_{A_l} = - \nelec \sigt \left(
\cle_{A_l} - \frac{2}{15} \zeta_{a_1 a_2} \delta_l^2 \right), \label{eq:47}\\
\dot{\clb}_{A_l} &+&\frac{4}{3}\Theta \clb_{A_l} + \frac{(l+3)(l-1)}{(l+1)^2}
\uD^b \clb_{b A_l} - \frac{l}{(2l+1)} \uD_{\langle a_l}\clb_{A_{l-1}\rangle}
\nonumber \\
&+& \frac{2}{(l+1)} \curl \cle_{A_l} = - \nelec \sigt \clb_{A_l}. \label{eq:48}
\end{eqnarray}
For the circular polarization, we find to $O(\epsilon)$
\begin{equation}
\dot{V}_{A_l} + \frac{4}{3}\Theta V_{A_l} - \frac{l}{(2l+1)}\uD_{\langle a_l}
V_{A_{l-1} \rangle} + \uD^b V_{b A_l} = -\nelec \sigt \left(V_{A_l}
-\frac{1}{2} V_{a_1} \delta_l^1 \right),
\label{eq:49}
\end{equation}
and for the intensity
\begin{eqnarray}
\dot{I}_{A_l} &+& \frac{4}{3}\Theta I_{A_l} + \uD^b I_{b A_l} -
\frac{l}{(2l+1)} \uD_{\langle a_l} I_{A_{l-1}\rangle}
+ \frac{4}{3} I A_{a_1}\delta_l^1 - \frac{8}{15} I \sigma_{a_1 a_2} \delta_l^2
\nonumber \\
&=& -\nelec \sigt \left(
I_{A_l} - I \delta_l^0 - \frac{4}{3} I v_{a_1} \delta_l^1 - \frac{2}{15}
\zeta_{a_1 a_2} \delta_l^2 \right),
\label{eq:50}
\end{eqnarray}
which extends the result in Refs.~\cite{LC-scalcmb,maartens99,gebbie99b}
to include the polarization dependence of Thomson scattering.
In Eq.~(\ref{eq:50}), $A_a \equiv \dot{u}_a$ is the acceleration of
$u^a$ and $\sigma_{ab} \equiv \uD_{\langle a} u_{b \rangle}$ is the shear.
Eqs.~(\ref{eq:47}--\ref{eq:50}) provide a complete description of radiative
transfer in almost-FRW models. They describe the evolution of the
energy-integrated intensity and polarization multipoles along the integral
curves of $u^a$. The equations are valid for any type of perturbation
(we have not decomposed the variables into their scalar, vector or tensor
parts), and for any (physical) choice of $u^a$. 
To close the equations it is necessary
to supplement them with the 1+3 covariant hydrodynamic equations,
e.g. Ref.~\cite{ellis98}, to determine the $O(\epsilon)$ kinematic variables
$A_a$, $v_a$, and $\sigma_{ab}$.

\section{Scalar perturbations}
\label{sec:scalar}

Up to this point our discussion has been quite general. Although we
specialised to almost-FRW models when discussing the multipole propagation
equations, we did not split
the perturbations into their constituent modes (scalar, vector, tensor etc.).
However, for detailed calculation with the linearised, almost-FRW equations,
it is convenient to exploit the linearity to break the problem into smaller
pieces. Our strategy follows Refs.~\cite{LC-scalcmb,gebbie99b,chall99b}:
we expand all $O(\epsilon)$ variables in
PSTF tensors derived from appropriate irreducible eigenfunctions of the
comoving Laplacian $S^2 \uD^a \uD_a$, where $S$ is the covariantly-defined
scale factor with $\dot{S}/S = \Theta/3$ and $\uD_a S=O(\epsilon)$.
Since the sources of anisotropy and polarization are at most rank-2
objects, there are non-vanishing contributions to the source terms only from
the scalar, vector and rank-2 tensor eigenfunctions. The different perturbation
types decouple at linear order, with each giving rise to a set of coupled,
first-order ordinary differential equations. We consider the scalar modes in
this section; tensor modes are treated in Sec.~\ref{sec:tens}. Since vorticity
dies away in expanding models in the absence of significant momentum
densities and anisotropic stresses~\cite{haw66}, we do not consider
vector modes. If required, e.g. in defect models~\cite{pen97}, vector modes
can be easily included in the formalism. The mode-expanded 1+3
covariant equations describing the perturbations in the matter components other
than the radiation, and the geometry, can be found in
Refs.~\cite{LC-scalcmb,gebbie99b,chall99b}, so we need only consider the
radiation here. Furthermore, we do not consider circular polarization any
further since it is not generated by Thomson scattering.

For scalar perturbations we expand in the scalar eigenfunctions
$Q^{(k)}$, where
\begin{equation}
S^2 \uD^a \uD_a Q^{(k)} = k^2 Q^{(k)}
\label{eq:51}
\end{equation}
at zero order. The $Q^{(k)}$ are constructed so that
$\dot{Q}^{(k)}=O(\epsilon)$.
The allowed eigenvalues $k^2$ depend on the spatial geometry of the background
model. Defining $\nu^2 = (k^2+K)/|K|$, where $6K/S^2$ is the curvature scalar
of the spatial sections in the background model, the regular, normalisable
eigenfunctions have $\nu \geq 0$ for open and flat models ($K \leq 0$). In
closed models $\nu$ is restricted to integer values
$\geq 1$~\cite{tomita82,abbott86}.
The mode with $\nu=1$ cannot be used to construct perturbations (its
projected gradient vanishes globally), while the mode with $\nu=2$ (which
can only represent isocurvature perturbations~\cite{bardeen80}) only
contributes to the CMB dipole. An explicit PSTF representation of the
scalar harmonics $Q^{(k)}$ is given in the appendix.

For the $l$-th multipoles of the radiation anisotropy and polarization we
expand in rank-$l$ PSTF tensors, $Q^{(k)}_{A_l}$, derived from
the scalar harmonics via
\begin{equation}
Q^{(k)}_{A_l} = \left(\frac{S}{k}\right)^{l}\uD_{\langle a_1}\dots
\uD_{a_l \rangle} Q^{(k)}.
\label{eq:52}
\end{equation}
The recursion relation for the $Q^{(k)}_{A_l}$,
\begin{equation}
Q^{(k)}_{A_l} = \frac{k}{S} \uD_{\langle a_l}Q^{(k)}_{A_{l-1}\rangle},
\label{eq:57}
\end{equation}
follows directly from Eq.~(\ref{eq:52}). The factor of $(S/k)^l$ in the
definition of the $Q^{(k)}_{A_l}$ ensures that $\dot{Q}^{(k)}_{A_l}=0$
at zero-order.
For $I_{A_l}$ and $\cle_{A_l}$ we write\footnote{The $I^{(l)}_k$ are related
to the $J^{(l)}_k$ of Ref.~\cite{LC-scalcmb} by $I^{(l)}_k = \alpha_l
J^{(l)}_k$.}
\begin{eqnarray}
I_{A_l} &=& I \sum_k \alpha_l{}^{-1} I^{(l)}_k Q^{(k)}_{A_l},
\qquad l \geq 1,\label{eq:53}\\
\cle_{A_l} &=& I \sum_k \alpha_l{}^{-1} \cle^{(l)}_k Q^{(k)}_{A_l},
\label{eq:54}
\end{eqnarray}
where we have defined $\alpha_l \equiv \prod_{n=1}^l \kappa_n$, with
\begin{equation}
\kappa_l \equiv [1-(l^2-1)K/k^2]^{1/2}, \qquad l \geq 1,
\label{eq:55}
\end{equation}
and $\alpha_0 =1$. The symbolic summation in Eqs.~(\ref{eq:53}) and
(\ref{eq:54}) denotes a sum over the harmonics (see Eq.~[\ref{eq:ap7}]), and
the mode coefficients, such as $I^{(l)}_k$, are $O(\epsilon)$ scalars
with $O(\epsilon^2)$ projected gradients. We need not consider the
magnetic polarization since it is not generated by scalar
modes~\cite{kamion97,seljak97}. To see this, we need the first-order identity
\begin{equation}
\curl \uD_{\langle a_{l+1}} S_{A_l \rangle} = \frac{l}{(l+1)}
\uD_{\langle a_{l+1}} \curl S_{A_l \rangle},
\label{eq:56}
\end{equation}
where $S_{A_l}$ is an $O(\epsilon)$, rank-$l$ PSTF tensor.
Applying this identity repeatedly to the right-hand side of
Eq.~(\ref{eq:57}) we see that the curl of $Q^{(k)}_{A_l}$ vanishes, as we
would expect for a scalar perturbation. It follows
that in linear theory $\curl \cle_{A_l} = 0$ for scalar perturbations, so
the inhomogeneous source terms in the propagation equation for
$\clb_{A_l}$ (Eq.~[\ref{eq:48}]) vanish. Since primordial polarization is
erased by Thomson scattering during tight coupling, scalar perturbations
do not support magnetic polarization.

The decomposition of $I(e^c)$ into angular multipoles $I_{A_l}$, and the
subsequent expansion in the $Q^{(k)}_{A_l}$, combine to give a normal mode
expansion which is equivalent to the Legendre tensor approach, first
introduced by Wilson~\cite{wilson83}. The advantage of handling the
angular and scalar harmonic decompositions separately is that the former
can be applied quite generally for an arbitrary cosmological model.
Furthermore, extending the normal mode expansions to the polarization and
tensor modes is then trivial once the angular decomposition has been
performed.

Substituting Eq.~(\ref{eq:53}) into Eq.~(\ref{eq:50}), and using the
zero-order identity~\cite{LC-scalcmb,gebbie99b}
\begin{equation}
\uD^{a_l} Q^{(k)}_{A_l} = \frac{k}{S}\frac{l}{(2l-1)}\left[1-(l^2-1)
\frac{K}{k^2}\right]Q^{(k)}_{A_{l-1}},
\label{eq:58}
\end{equation}
gives the Boltzmann hierarchy for the $I^{(l)}_k$:
\begin{eqnarray}
\dot{I}^{(l)}_k &+& \frac{k}{S}\left[\frac{(l+1)}{(2l+1)}\kappa_{l+1}
I^{(l+1)}_k - \frac{l}{(2l+1)}\kappa_l I^{(l-1)}_k \right]
+ \frac{4}{3}\left(\frac{k}{S}\clz_k -\Theta A_k \right)\delta_l^0 \nonumber\\
&+& \frac{4}{3}\frac{k}{S} A_k \delta_l^1
- \frac{8}{15}\frac{k}{S} \kappa_2 \sigma_k \delta_l^2
= -\nelec \sigt\left(I^{(l)}_k - \delta_l^0 I^{(0)}_k -
\frac{4}{3}\delta_l^{(1)}v_k - \frac{2}{15} \zeta_k \delta_l^2 \right).
\label{eq:59}
\end{eqnarray}
It should be noted that $I^{(0)}_k$ refers to the projected gradient of
intensity monopole $I$,
\begin{equation}
\frac{\uD_a I}{I} = \sum_k \frac{k}{S} I^{(0)}_k Q^{(k)}_a,
\label{eq:60}
\end{equation}
which is gauge-invariant. To obtain Eq.~(\ref{eq:59}) for $l=0$, take
the projected gradient of Eq.~(\ref{eq:50}) for $l=0$ and commute
the derivatives. This gives rise to the source term $\clz_k$, where
\begin{equation}
\uD_a \Theta = \sum_k \left(\frac{k}{S}\right)^2 \clz_k Q^{(k)}_a.
\label{eq:61}
\end{equation}
The other kinematic source terms come from the acceleration $A_a$, the
the baryon relative velocity $v_a$, and the shear $\sigma_{ab}$, with
\begin{eqnarray}
A_a &=& \sum_k \frac{k}{S} A_k Q^{(k)}_a, \label{eq:62}\\
v_a &=& \sum_k v_k Q^{(k)}_a, \label{eq:63}\\
\sigma_{ab} &=& \sum_k \frac{k}{S} \sigma_k Q^{(k)}_{ab}.\label{eq:64}
\end{eqnarray}
The term describing the dependence of Thomson scattering on the anisotropy
and polarization is $\zeta_k = 3I^{(2)}_k/4 + 9\cle^{(2)}_k/2$.
Eq.~(\ref{eq:59}) extends the 1+3 covariant results of
Refs.~\cite{LC-scalcmb,gebbie99b} by including this polarization source term.

For the electric polarization we use Eq.~(\ref{eq:54}) in Eq.~(\ref{eq:47})
to find
\begin{equation}
\dot{\cle}^{(l)}_k + \frac{k}{S}\left[\frac{(l+3)(l-1)}{(2l+1)(l+1)}
\kappa_{l+1}\cle^{(l+1)}_k - \frac{l}{(2l+1)}\kappa_l \cle^{(l-1)}_k \right]
= -\nelec \sigt\left(\cle^{(l)}_k - \frac{2}{15}\zeta_k \delta_l^2\right),
\label{eq:65}
\end{equation}
which is the 1+3 covariant analogue of equivalent results
in Refs.~\cite{hu98,seljak98}. Eqs.~(\ref{eq:59}) and (\ref{eq:65}) can be
integrated with 1+3 covariant hydrodynamic
equations~\cite{LC-scalcmb,gebbie99b} to determine the anisotropy and
polarization from scalar modes.

To obtain formal integral solutions to the Boltzmann hierarchies we note
that the homogeneous form of Eq.~(\ref{eq:59}), obtained by setting
$\nelec=0$ and removing the kinematic source terms, is solved by
$\Phi_l^\nu(x)$, where $x=\sqrt{|K|}
(\eta_R-\eta)$, with $\eta_R$ the conformal time at our current
position $R$. Here $\Phi_l^\nu(x)$ are the ultra-spherical Bessel
functions~\cite{abbott86}. The full solution of Eq.~(\ref{eq:59}) follows
from Green's method:
\begin{eqnarray}
I^{(l)}_k &=& 4 \int^{t_R}\ud t\, e^{-\tau}\Bigg\{\left(\frac{k}{S}
\sigma_k + \frac{1}{4} \nelec \sigt \kappa_2{}^{-1}\zeta_k\right)
\left[\frac{1}{3} \Phi_l^\nu(x) + \frac{1}{(\nu^2+1)} \frac{\ud^2}{\ud x^2}
\Phi_l^\nu(x)\right] \nonumber \\
&& - \left(\frac{k}{S}A_k - \nelec \sigt v_k \right)\frac{1}{\sqrt{\nu^2+1}}
\frac{\ud}{\ud x}\Phi_l^\nu(x) - \left[\frac{1}{3}\left(\frac{k}{S}\clz_k
-\Theta A_k\right)-\frac{1}{4}\nelec\sigt I^{(0)}_k \right] \Phi_l^\nu(x)
\Bigg\},
\label{eq:66}
\end{eqnarray}
where $\tau$ is the zero-order optical depth back to $x$
(see Eq.~[\ref{eq:42}]).
The geometric factors $\Phi_l^\nu/3 + (\nu^2+1)^{-1}\ud^2 \Phi_l^\nu/\ud x^2$
and $(\nu^2+1)^{-1/2} \ud \Phi_l^\nu/\ud x$ arise from the projections
of $Q^{(k)}_{ab}e^a e^b$ and $Q^{(k)}_a e^a$ respectively,
at $x$ back along the line of sight.

For the polarization we note that the homogeneous part of Eq.~(\ref{eq:65})
is solved by $l(l-1)\Phi_l^\nu(x)/\sinh^2\! x$. The full solution is
\begin{equation}
\cle^{(l)}_k = \frac{l(l-1)}{(\nu^2+1)} \int^{t_R} \ud t\, \nelec
\sigt e^{-\tau} \kappa_2{}^{-1} \zeta_k \frac{\Phi_l^\nu(x)}{\sinh^2 \! x}.
\label{eq:67}
\end{equation}
Eqs.~(\ref{eq:66}) and (\ref{eq:67}) are valid in an open universe. For closed
models one should replace the hyperbolic functions by their trigonometric
counterparts, and $\nu^2+n$ by $\nu^2-n$ where $n$ is an integer.

To characterise the initial amplitude of the mode we introduce a set of
random variables $\phi_k$ with the covariance matrix
\begin{equation}
\langle \phi_k \phi_{k'} \rangle = \frac{1}{|K|^{3/2}}
\frac{\clp_\phi(\nu)}{\nu(\nu^2+1)} \delta_{kk'},
\label{eq:68}
\end{equation}
appropriate to a statistically isotropic and homogeneous ensemble. The
$\delta_{kk'}$ is defined by $\sum_k \Delta_k \delta_{kk'} = \Delta_{k'}$
for any mode functions $\Delta_k$ (see appendix). We follow the conventions
of Lyth \& Woszczyna~\cite{lyth95}, so that the scalar field $\sum_k
\phi_k Q^{(k)}$ has a scale-invariant spectrum for $\clp_\phi(\nu)=
\text{constant}$. We write the radiation mode coefficients in terms of
the $\phi_k$ and transfer functions $T^{(l)}_I(\nu)$ and $T^{(l)}_\cle(\nu)$:
\begin{equation}
I^{(l)}_k = T^{(l)}_I(\nu) \phi_k, \qquad
\cle^{(l)}_k = \frac{M_l}{\sqrt{2}} T^{(l)}_\cle(\nu) \phi_k,
\label{eq:69}
\end{equation}
where the factor $M_l/\sqrt{2}$ is for later convenience.
To compute the anisotropy and polarization power spectra we use
Eqs.~(\ref{eq:53}) and (\ref{eq:54}) in Eqs.~(\ref{eq:16}) and (\ref{eq:19})
to find
\begin{equation}
C_l^{XY} = \frac{1}{16} \int_0^\infty \frac{\nu \ud \nu}{(\nu^2+1)}
T^{(l)}_X(\nu) T^{(l)}_Y(\nu) \clp_\phi(\nu).
\label{eq:70}
\end{equation}
Here, $X$ and $Y$ represent either of $I$ or $\cle$.
In a closed universe the integral over $\nu$ in Eq.~(\ref{eq:70})
should be replaced by a sum over integral $\nu \geq l+1$.
To derive Eq.~(\ref{eq:70}) we have made use of the result
\begin{equation}
\sum_k f(\nu) Q^{(k)}_{A_l} Q^{(k)B_{l'}} =
\frac{1}{(4\pi)^2}\Delta_l h^{\langle B_l \rangle}_{\langle A_l \rangle}
\delta_l^{l'} \int_0^\infty \ud \nu\, |K|^{3/2} \nu^2 f(\nu) \alpha_l{}^2,
\label{eq:71}
\end{equation}
for any scalar function $f(\nu)$, which follows from Eqs.~(\ref{eq:ap6})
and (\ref{eq:ap9}) in the appendix.

\section{Tensor perturbations}
\label{sec:tens}

For tensor perturbations we follow the procedure in Ref.~\cite{chall99b}
and expand in rank-2 PSTF tensor eigenfunctions $Q^{(k)}_{ab}$ of the comoving
Laplacian:
\begin{equation}
S^2 \uD^c \uD_c Q^{(k)}_{ab} = k^2 Q^{(k)}_{ab}.
\label{eq:72}
\end{equation}
The tensor harmonics are transverse, $D^a Q^{(k)}_{ab}=0$, and constant along
the integral curves of $u^a$, $\dot{Q}^{(k)}_{ab}=0$, at
zero-order. For tensor modes we define $\nu^2=(k^2+3K)/|K|$. The regular,
normalisable eigenmodes have $\nu \geq 0$ for flat and open models, while for
closed models $\nu$ is an integer $\geq 3$. Explicit forms for the tensor
harmonics in the PSTF representation are given in the appendix; see also
Ref.~\cite{chall99b}. The tensor harmonics can be classified as having
electric or magnetic parity. We continue to denote the electric parity
harmonics by $Q^{(k)}_{ab}$, but we use an overbar to distinguish the magnetic
parity harmonics: $\bar{Q}^{(k)}_{ab}$. The electric and magnetic parity
harmonics are related through the curl operation (see Eqs.~[\ref{eq:ap19}]
and [\ref{eq:ap20}]).

Following our treatment of scalar perturbations, we form rank-$l$ PSTF tensors
$Q^{(k)}_{A_l}$ and $\bar{Q}^{(k)}_{A_l}$ from the electric and
magnetic parity tensor harmonics respectively. We define
\begin{equation}
Q^{(k)}_{A_l} = \left(\frac{S}{k}\right)^{l-2} \uD_{\langle a_1}
\dots \uD_{a_{l-2}} Q^{(k)}_{a_{l-1} a_l \rangle},
\label{eq:73}
\end{equation}
with an equivalent definition for the magnetic parity harmonics. The
rank-$l$ tensors satisfy $\dot{Q}^{(k)}_{A_l}=0$ at zero-order. We
expand the $l$-th multipoles of the intensity and linear polarization
as
\begin{eqnarray}
I_{A_l} &=& I \sum_k \beta_l{}^{-1} (I^{(l)}_k Q^{(k)}_{A_l} +
\bar{I}^{(l)}_k \bar{Q}^{(k)}_{A_l}), \qquad l\geq 2,\label{eq:74}\\
\cle_{A_l} &=& I \sum_k \beta_l{}^{-1} (\cle^{(l)}_k Q^{(k)}_{A_l} +
\bar{\cle}^{(l)}_k \bar{Q}^{(k)}_{A_l}), \label{eq:75}\\
\clb_{A_l} &=& I \sum_k \beta_l{}^{-1} (\clb^{(l)}_k Q^{(k)}_{A_l} +
\bar{\clb}^{(l)}_k \bar{Q}^{(k)}_{A_l}). \label{eq:76}
\end{eqnarray}
For tensor modes we have defined $\beta_l \equiv \prod_{n=2}^l \kappa_n$,
and
\begin{equation}
\kappa_l \equiv [1-(l^2-3)K/k^2]^{1/2}, \qquad l\geq 2.
\label{eq:77}
\end{equation}
It is necessary to include the magnetic polarization for tensor modes since
the source to the magnetic hierarchy does not vanish:
$\curl \cle_{A_l} \neq 0$.

The $Q^{(k)}_{A_l}$ constructed from the tensor harmonics satisfy the same
recursion relation, Eq.~(\ref{eq:57}), as their scalar counterparts.
However, their projected divergences evaluate to~\cite{chall99b}
\begin{equation}
\uD^{a_l} Q^{(k)}_{A_l} = \frac{k}{S} \frac{(l^2-4)}{l(2l-1)}
\left[1-(l^2-3)\frac{K}{k^2}\right] Q^{(k)}_{A_{l-1}}.
\label{eq:78}
\end{equation}
An equivalent relation holds for the $\bar{Q}^{(k)}_{A_l}$. Using these
results in Eq.~(\ref{eq:50}) we find
\begin{eqnarray}
\dot{I}^{(l)}_k &+& \frac{k}{S}\left[\frac{(l+3)(l-1)}{(l+1)(2l+1)}
\kappa_{l+1} I^{(l+1)}_k - \frac{l}{(2l+1)} \kappa_l I^{(l-1)}_k \right]
- \frac{8}{15} \frac{k}{S} \kappa_2 \sigma_k \delta_l^2 \nonumber \\
&=& - \nelec \sigt \left(I^{(l)}_k - \frac{2}{15} \zeta_k \delta_l^2 \right),
\label{eq:79}
\end{eqnarray}
where $\zeta_k = 3I^{(2)}_k/4 + 9\cle^{(2)}_k/2$, and $\sigma_k$
is the tensor mode coefficient for the shear,
\begin{equation}
\sigma_{ab} =\sum_k \frac{k}{S}(\sigma_k Q^{(k)}_{ab} + \bar{\sigma}_k
\bar{Q}^{(k)}_{ab}).
\label{eq:80}
\end{equation}
Eq.~(\ref{eq:79}) extends the 1+3 covariant result in Ref.~\cite{chall99b}
to include the polarization source term. For the polarization hierarchies
we need the result
\begin{equation}
\curl Q^{(k)}_{A_l} = \frac{2}{l}\frac{k}{S} \sqrt{1+\frac{3K}{k^2}}
\bar{Q}^{(k)}_{A_l},
\label{eq:81}
\end{equation}
which follows from repeated application of Eq.~(\ref{eq:56}), and finally
Eq.~(\ref{eq:ap19}) to express $\curl Q^{(k)}_{a_{l-1} a_l}$ in terms of
$\bar{Q}^{(k)}_{a_{l-1} a_l}$. For the electric polarization, Eq.~(\ref{eq:47})
gives
\begin{eqnarray}
\dot{\cle}^{(l)}_k &+& \frac{k}{S}\left[\frac{(l+3)^2(l-1)^2}{(2l+1)(l+1)^3}
\kappa_{l+1} \cle^{(l+1)}_k - \frac{l}{(2l+1)} \kappa_l \cle^{(l-1)}_k
\right] - \frac{4}{l(l+1)} \frac{k}{S} \sqrt{1+\frac{3K}{k^2}}
\bar{\clb}^{(l)}_k \nonumber \\
&=& -\nelec \sigt\left(\cle^{(l)}_k - \frac{2}{15} \zeta_k \delta_l^2 \right).
\label{eq:82}
\end{eqnarray}
Eq.~(\ref{eq:48}) gives the corresponding equation for the magnetic
polarization:
\begin{eqnarray}
\dot{\clb}^{(l)}_k &+& \frac{k}{S}\left[\frac{(l+3)^2(l-1)^2}{(2l+1)(l+1)^3}
\kappa_{l+1} \clb^{(l+1)}_k - \frac{l}{(2l+1)} \kappa_l \clb^{(l-1)}_k
\right] + \frac{4}{l(l+1)} \frac{k}{S} \sqrt{1+\frac{3K}{k^2}}
\bar{\cle}^{(l)}_k \nonumber \\
&=& -\nelec \sigt \clb^{(l)}_k.
\label{eq:83}
\end{eqnarray}
Equivalent equations hold for the barred variables $\bar{\cle}^{(l)}_k$
and $\bar{\clb}^{(l)}_k$. Note how power is transferred between
$\bar{\clb}^{(l)}_k$ and $\cle^{(l)}_k$ due to the curl coupling
between the electric and magnetic multipoles in Eqs.~(\ref{eq:47}) and
(\ref{eq:48}).

The integral solution for the tensor contribution to the intensity anisotropy
in a general almost-FRW model was given in Ref.~\cite{chall99b}. Including
polarization modifies the scattering source term, so that the integral solution
becomes
\begin{equation}
I^{(l)}_k = \frac{4l(l-1)}{[(\nu^2+1)(\nu^2+3)]^{1/2}}\int^{t_R}
\ud t\, e^{-\tau} \left(\frac{k}{S}\sigma_k + \frac{1}{4}\nelec\sigt
\kappa_2{}^{-1} \zeta_k \right) \frac{\Phi^\nu_l(x)}{\sinh^2\! x},
\label{eq:84}
\end{equation}
in an open universe.
The geometric factor $l(l-1)\Phi^\nu_l(x)/\sinh^2\! x$ follows from
the projection of $Q^{(k)}_{ab} e^a e^b$ at $x$ back along the line of
sight.

To solve the coupled equations (\ref{eq:82}) and (\ref{eq:83}) for the
linear polarization it is simplest to return to the integral solution for
the tetrad components of $\clp_{ab}(e^c)$, given as Eq.~(\ref{eq:41}).
In linearised form, we have
\begin{equation}
I^{-1} \clp_{ab}(e^c)(e_i)^a (e_j)^b |_R = \frac{1}{4\pi}
\int^{t_R} \ud t\, \nelec \sigt e^{-\tau}I^{-1}[\zeta_{ab}]^{\TT}
(e_i)^a (e_j)^b.
\label{eq:85}
\end{equation}
Substituting the tensor mode expansion of $\zeta_{ab}$ gives
terms in the integral like $[Q^{(k)}_{ab}]^{\TT}(e_i)^a (e_j)^b$ from the
electric parity harmonics, and similar terms from the harmonics with magnetic
parity. Using the representation of the tensor harmonics given in the
appendix, we find that
\begin{equation}
[Q^{(k)}_{ab}]^{\TT}(e_i)^a (e_j)^b = T_3(x)
[\clq^{(LM)}_{ab C_{L-2}} e^{C_{L-2}}]^{\TT} (e_i)^a (e_j)^b,
\label{eq:86}
\end{equation}
where $T_3(x)$ is a $\nu$ and $L$-dependent function given by
Eq.~(\ref{eq:ap13}). The superscript $(k)$ on the tensor harmonics represents
the collection $\nu$, $L$, and $M$, where $L$ and $M$ describe the orbital
angular momentum of the harmonic; see the appendix for details.
The $\clq^{(LM)}_{A_L}$ are rank-$L$ PSTF tensor fields, which are introduced
in the appendix. They encode the $2L+1$ degrees of freedom in the tensor
harmonics with $L$ units of orbital angular momentum.
Using the transport properties of $\clq^{(LM)}_{A_L}$,
Eq.~(\ref{eq:ap3}), and of the tetrad vectors $(e_i)^a$, Eq.~(\ref{eq:32}),
it is straightforward to show that to zero-order
\begin{equation}
(u^a + e^a)\nabla_a \{[\clq^{(LM)}_{ab C_{L-2}} e^{C_{L-2}}]^{\TT}
(e_i)^a (e_j)^b \} = 0,
\label{eq:87}
\end{equation}
so that $[\clq^{(LM)}_{ab C_{L-2}} e^{C_{L-2}}]^{\TT} (e_i)^a (e_j)^b$ is
constant along the line of sight. For the magnetic parity harmonics,
the representation in the appendix gives
\begin{equation}
[\bar{Q}^{(k)}_{ab}]^{\TT} (e_i)^a (e_j)^b = \bar{T}_1(x)
[e_{c_L} \epsilon^{c_L c_{L-1}}{}_{(a}\clq^{(LM)}_{b)C_{L-1}}e^{C_{L-2}}]^{\TT}
(e_i)^a (e_j)^b, 
\label{eq:88}
\end{equation}
where $\bar{T}_1(x)$ is given by Eq.~(\ref{eq:ap18}). Working to zero-order,
we find that
\begin{equation}
(u^a + e^a)\nabla_a \{[e_{c_L} \epsilon^{c_L c_{L-1}}{}_{(a}
\clq^{(LM)}_{b)C_{L-1}}e^{C_{L-2}}]^{\TT} (e_i)^a (e_j)^b \} = 0,
\label{eq:89}
\end{equation}
so that $[e_{c_L} \epsilon^{c_L c_{L-1}}{}_{(a}
\clq^{(LM)}_{b)C_{L-1}}e^{C_{L-2}}]^{\TT} (e_i)^a (e_j)^b$ is also constant
along the line of sight at zero-order. With these results, Eq.~(\ref{eq:85})
reduces to
\begin{eqnarray}
I^{-1} \clp_{ab}(e^c)|_R &=& \frac{1}{4\pi}
\sum_k [\clq^{(LM)}_{ab C_{L-2}} e^{C_{L-2}}]^{\TT}
\int^{t_R} \ud t\, \nelec \sigt e^{-\tau} \kappa_2{}^{-1} \zeta_k T_3(x)
\nonumber \\
&& + \frac{1}{4\pi} \sum_k [e_{c_L} \epsilon^{c_L c_{L-1}}{}_{(a}
\clq^{(LM)}_{b)C_{L-1}} e^{C_{L-2}}]^{\TT} \int^{t_R} \ud t\, \nelec \sigt
e^{-\tau} \kappa_2{}^{-1} \bar{\zeta}_k \bar{T}_1(x).
\label{eq:90}
\end{eqnarray}
Extracting the $l$-th electric and magnetic parity multipoles from
this equation, we find
\begin{eqnarray}
\cle_{A_{l}}|_R &=& \frac{1}{4\pi} \sum_k \delta_{lL} \Delta_l
\clq^{(LM)}_{A_L}|_R
\int^{t_R} \ud t\, \nelec \sigt e^{-\tau} \kappa_2{}^{-1} \zeta_k T_3(x),
\label{eq:91} \\
\clb_{A_{l}}|_R &=& \frac{1}{4\pi} \sum_k \delta_{lL} \Delta_l
\clq^{(LM)}_{A_L}|_R
\int^{t_R} \ud t\, \nelec \sigt e^{-\tau} \kappa_2{}^{-1} \bar{\zeta}_k
\bar{T}_1(x).
\label{eq:92}
\end{eqnarray}
To determine the harmonic coefficients, such as $\cle^{(l)}_k$, at $t_R$
we note that $\bar{Q}^{(k)}_{A_{l}}=0$ at $R$, so that only the electric
parity harmonics contribute there. We can use Eq.~(\ref{eq:ap15}) to
substitute for $\delta_{lL}\clq^{(LM)}_{A_L}$ at each $\nu$ in terms of
$Q^{(k)}_{A_{l}}|_R$ in Eqs.~(\ref{eq:91}) and (\ref{eq:92}).
Comparing with Eqs.~(\ref{eq:75})
and (\ref{eq:76}), we can read off the integral solutions
\begin{eqnarray}
\cle^{(l)}_k &=& M_l \frac{\nu}{\sqrt{\nu^2+3}}
\int^{t_R} \ud t\, \nelec \sigt e^{-\tau} \kappa_2{}^{-1} \zeta_k T_3(x),
\label{eq:93} \\
\clb^{(l)}_k &=& M_l \frac{\nu}{\sqrt{\nu^2+3}}
\int^{t_R} \ud t\, \nelec \sigt e^{-\tau} \kappa_2{}^{-1} \bar{\zeta}_k
\bar{T}_1(x).
\label{eq:94}
\end{eqnarray}
Evaluating the functions $T_3(x)$ and $\bar{T}_1(x)$, we can write the
integral solutions in the form
\begin{eqnarray}
\cle^{(l)}_k &=& \frac{M_l{}^2}{2[(\nu^2+1)(\nu^2+3)]^{1/2}}
\int^{t_R} \ud t\, \nelec \sigt e^{-\tau} \kappa_2{}^{-1} \zeta_k\phi_l^\nu(x),
\label{eq:95} \\
\clb^{(l)}_k &=& \frac{M_l{}^2}{2[(\nu^2+1)(\nu^2+3)]^{1/2}}
\int^{t_R} \ud t\, \nelec \sigt e^{-\tau} \kappa_2{}^{-1} \bar{\zeta}_k
\psi_l^\nu(x),
\label{eq:96}
\end{eqnarray}
where the geometric terms $\phi_l^\nu(x)$ and $\psi^\nu_l(x)$ are
\begin{eqnarray}
\phi^\nu_l(x) &=& \frac{\ud^2}{\ud x^2}\Phi^\nu_l(x) + 4\coth\! x
\frac{\ud}{\ud x}\Phi^\nu_l(x) - (\nu^2-1-2\coth^2 x) \Phi^\nu_l(x),
\label{eq:97} \\
\psi^\nu_l(x) &=& -2\nu \left[ \frac{\ud}{\ud x}\Phi^\nu_l(x)
+ 2\coth\! x \Phi^\nu_l(x) \right]
\label{eq:98}
\end{eqnarray}
in an open universe. In closed models we replace $\nu^2 +n$ by $\nu^2-n$, and
the hyperbolic functions by their trigonometric counterparts (as for scalar
perturbations). It is straightforward to verify that the integral
solutions, Eqs.~(\ref{eq:95}) and (\ref{eq:96}), satisfy the multipole
equations~(\ref{eq:82}) and (\ref{eq:83}). The solutions for
$\bar{\cle}^{(l)}_k$ and $\bar{\clb}^{(l)}_k$ are the same as those for
$\cle^{(l)}_k$ and $\clb^{(l)}_k$ but with the replacement $\zeta_k
\leftrightarrow \bar{\zeta}_k$ for the source function. Although the
magnetic parity tensor harmonics do not contribute at $R$, they do contribute
to the anisotropy and polarization at points not on the integral curve of
$u^a$ that passes through $R$. The integral solutions given here agree with
those of Ref.~\cite{hu98} which were derived with the total angular momentum
method.

For tensor perturbations we characterise the initial amplitude of the modes
by introducing random variables $\phi_k$ and $\bar{\phi}_k$ with the
covariance structure
\begin{equation}
\langle \phi_k \phi_{k'} \rangle = \langle \bar{\phi}_k \bar{\phi}_{k'}
\rangle = \frac{1}{|K|^{3/2}}\frac{\clp_{\phi}(\nu)}{\nu(\nu^2+1)}
\delta_{k k'}, \qquad \langle \phi_k \bar{\phi}_{k'} \rangle = 0,
\label{eq:99} 
\end{equation}
appropriate to statistical homogeneity and isotropy. Note that this form
for the covariance structure forbids any cross correlations between the
$\clb_{A_l}$ and $\cle_{A_l}$ or $I_{A_l}$. In
Ref.~\cite{chall99b} we took $\phi_k = (\nu^2+3)E_k / (\nu^2+1)$
where $E_k$ is the initial amplitude of the dimensionless mode coefficient
representing the electric part of the Weyl tensor. In this case, the
minimal scale-invariant prediction of one bubble open inflation
is $\clp_\phi=\tanh (\pi \nu/2)$~\cite{bucher97}. We write the mode
coefficients for the intensity and the polarization in the form
\begin{equation}
I^{(l)}_k = T^{(l)}_I(\nu) \phi_k, \qquad
\cle^{(l)}_k = \frac{M_l}{\sqrt{2}} T^{(l)}_\cle(\nu) \phi_k, \qquad
\clb^{(l)}_k = \frac{M_l}{\sqrt{2}} T^{(l)}_\clb(\nu) \bar{\phi}_k.
\label{eq:100}
\end{equation}
Note that $\clb^{(l)}_k$ is proportional to $\bar{\phi}_k$ since the source
to $\clb^{(l)}_k$ is $\bar{\zeta}_k$ rather than $\zeta_k$; see
Eq.~(\ref{eq:96}). We can now compute the anisotropy and polarization
polarization power spectra for tensor modes from Eqs.~(\ref{eq:16})
and (\ref{eq:19}). The result is
\begin{equation}
C_l^{XY} = \frac{1}{16} \frac{(l+2)(l+1)}{2l(l-1)} \int_0^\infty
\frac{\nu \ud\nu}{(\nu^2+1)} \frac{(\nu^2+3)}{\nu^2} T^{(l)}_X(\nu)
T^{(l)}_Y (\nu) \clp_\phi(\nu),
\label{eq:101}
\end{equation}
where $XY$ is equal to $II$, $\cle\cle$, $\clb\clb$, or $I\cle$.
In closed models the integral over $\nu$ is replaced by a discrete sum
over integral $\nu \geq l+1$. In deriving Eq.~(\ref{eq:101}) we used the
following result:
\begin{equation}
\sum_k f(\nu) (Q^{(k)}_{A_l} Q^{(k)B_{l'}} + \bar{Q}^{(k)}_{A_l}
\bar{Q}^{(k)B_{l'}}) = \frac{1}{(4\pi)^2} M_l{}^{-2}
\Delta_l h^{\langle B_l \rangle}_{\langle A_l \rangle}
\delta_l^{l'} \int_0^\infty \ud \nu |K|^{3/2} (\nu^2+3) f(\nu) \beta_l{}^2, 
\label{eq:102}
\end{equation}
for any scalar function $f(\nu)$. This result is easily verified at $R$ by
using Eq.~(\ref{eq:ap15}) and the fact that $\bar{Q}^{(k)}_{A_l}|_R=0$.

\section{Discussion}
\label{sec:disc}

Most of the modern literature on the polarization of the CMB
employs representations based on either a coordinate representation of the
tensor spherical harmonics, introduced to CMB research in Ref.~\cite{kamion97},
or the Newman-Penrose spin-weight 2 harmonics (e.g. Ref.~\cite{goldberg67}),
first used for describing polarization in Ref.~\cite{seljak97}. Given the
widespread use of these two formalisms, it is worthwhile outlining
their relation to the PSTF representation adopted here. We begin by introducing
an orthonormal triad of projected vectors $\{(\gamma_1)^a, (\gamma_2)^a,
(\gamma_3)^a\}$ at the observer's location. At this point, the triad is
used to define angular coordinates $\{\theta,\phi\}$ which cover the
two-dimensional manifold of unit projected vectors $\{n^a\}$.
The scalar spherical harmonics $Y_{(lm)}(\theta,\phi)=Y_{(lm)}(n^a)$
can be represented by complex PSTF tensors $\cly^{(lm)}_{A_l}$, where
\begin{equation}
\cly^{(lm)}_{A_l} = \Delta_l{}^{-1} \int \ud \Omega\, Y_{(lm)}(n^c)
n_{\langle A_l \rangle},
\label{eq:103}
\end{equation}
so that $Y_{(lm)}(n^c)=\cly^{(lm)}_{A_l} n^{A_l}$. Using the
spherical harmonic addition theorem, one can show that the $\cly^{(lm)}_{A_l}$
satisfy the orthogonality relations
\begin{eqnarray}
\cly^{(lm)\ast}_{A_l} \cly^{(lm')A_l} &=& \Delta_l{}^{-1} \delta_{mm'},
\label{eq:104} \\
\sum_{m=-l}^l \cly^{(lm)\ast}_{A_l} \cly^{(lm)B_l} &=& \Delta_l{}^{-1}
h_{\langle A_l \rangle}^{\langle B_l \rangle},
\label{eq:105}
\end{eqnarray}
which are useful for the discussion below.

If we compare the PSTF expansion of the temperature anisotropy,
Eq.~(\ref{eq:17}), with Eq.~(2) of Ref.~\cite{kamion97} we can read off the
relation between the PSTF multipoles $I_{A_l}$ and the complex scalar
multipole coefficients $a_{(lm)}^{\text{T}}$ of the anisotropy:
\begin{equation}
\sum_{m=-l}^l a_{(lm)}^{\text{T}} \cly^{(lm)}_{A_l} =
\frac{\pi}{I \Delta_l} (-1)^l I_{A_l}.
\label{eq:106}
\end{equation}
The factor of $(-1)^l$ in this equation arises because in Ref.~\cite{kamion97}
the unit projected vector $n^a$ represents a direction on the sky, rather
than the photon propagation direction, so that $e^a=-n^a$. Using
Eqs.~(\ref{eq:105}) and~(\ref{eq:106}) in the expression for the temperature
power spectrum, Eq.~(\ref{eq:16}), we find that
\begin{equation}
\langle a_{(lm)}^{\text{T}\ast} a_{(l'm')}^{\text{T}}\rangle =
C_l^{II} \delta_{ll'}\delta_{mm'}.
\label{eq:107}
\end{equation}
It follows that the $C_l^{\text{T}}$ of Ref.~\cite{kamion97} coincides with
the temperature power spectrum defined here.

For the linear polarization we need to relate the PSTF representation of the
TT tensor spherical harmonics to the coordinate representation derived by
covariant differentiation of the $Y^{(lm)}(n^a)$. Taking
covariant derivatives on the sphere of the PSTF representation of
the $Y^{(lm)}(n^a)$, and using the conventions for the G(radient) and C(url)
tensor spherical harmonics of Ref.~\cite{kamion97}, we find that
\begin{eqnarray}
Y_{(lm)ab}^{\text{G}}(n^c)&=&M_l[\cly^{(lm)}_{ab C_{l-2}}n^{C_{l-2}}]^{\TT},
\label{eq:108} \\
Y_{(lm)ab}^{\text{C}}(n^c) &=& M_l
[n_{d_1}\epsilon^{d_1 d_2}{}_{(a}\cly_{b)d_2 C_{l-2}}^{(lm)}n^{C_{l-2}}]^{\TT}.
\label{eq:109}
\end{eqnarray}
Integrating Eq.~(\ref{eq:11}) over energy and dividing by $\pi/I$ expresses
the linear polarization in dimensionless temperature units.
Comparing with Eq.~(2) of
Ref.~\cite{kamion97}, we find the following relations between the
coordinate-dependent, scalar-valued multipole coefficients
$a_{(lm)}^{\text{G}}$ and $a_{(lm)}^{\text{C}}$ defined there, and the
PSTF multipoles $\cle_{A_l}$ and $\clb_{A_l}$ employed here:
\begin{eqnarray}
\frac{\pi}{I\Delta_l}(-1)^l \cle_{A_l} = M_l \sum_{m=-l}^l
a_{(lm)}^{\text{G}} \cly^{(lm)}_{A_l}, \label{eq:110}\\
\frac{\pi}{I\Delta_l}(-1)^{l-1} \clb_{A_l} = M_l \sum_{m=-l}^l
a_{(lm)}^{\text{C}} \cly^{(lm)}_{A_l}. \label{eq:111}
\end{eqnarray}

Using these results in the definitions of the polarization power spectra,
e.g. Eq.~(\ref{eq:19}), we find the non-vanishing covariance structure
\begin{eqnarray}
2 \langle a_{(lm)}^{\text{G}} a_{(l'm')}^{\text{G}\ast} \rangle &=&
C_l^{\cle\cle} \delta_{ll'} \delta_{mm'}, \label{eq:112} \\
2 \langle a_{(lm)}^{\text{C}} a_{(l'm')}^{\text{C}\ast} \rangle &=&
C_l^{\clb\clb} \delta_{ll'} \delta_{mm'},\label{eq:113} \\
\sqrt{2} \langle a_{(lm)}^{\text{T}} a_{(l'm')}^{\text{G}\ast} \rangle &=&
C_l^{I\cle} \delta_{ll'} \delta_{mm'}. \label{eq:114}
\end{eqnarray}
in a parity-symmetric ensemble. Comparing with Eq.~(8) of Ref.~\cite{kamion97},
we find that $2 C_l^{\text{G}} = C_l^{\cle\cle}$ and
$2 C_l^{\text{C}} = C_l^{\clb\clb}$, while $\sqrt{2}C_l^{\text{TG}}=
C_l^{I\cle}$.

To compare the PSTF representation of the linear polarization with the
spin-weight 2 representation of Ref.~\cite{seljak97}, it is simplest to express
the coordinate representation of the tensor spherical harmonics of
Ref.~\cite{kamion97} in terms of the spin-weight 2 spherical harmonics.
Noting that the Stokes parameters
in Ref.~\cite{seljak97} are expressed on a right-handed basis with the 3-axis
opposite to the direction of propagation (which flips the sign of $U$),
we find that for the $a_{E,lm}$ and $a_{B,lm}$ multipole coefficients of
Ref.~\cite{seljak97}, $a_{E,lm}=-\sqrt{2}a^{\text{G}}_{(lm)}$ and
$a_{B,lm}=-\sqrt{2}a^{\text{C}}_{(lm)}$, whereas the temperature multipoles
are equal. It follows that the polarization power spectra of
Ref.~\cite{seljak97} are related to those defined here by
$C_{El}=C_l^{\cle\cle}$, $C_{Bl}=C_l^{\clb\clb}$, and $C_{Cl}=-C_l^{I\cle}$.

\section{Conclusion}
\label{sec:conc}

We have introduced a new multipole formalism for describing polarized
radiation on the sky. In this approach the polarization tensor is expanded
in the PSTF representation of the transverse-traceless tensor spherical
harmonics~\cite{thorne80}. The PSTF representation is particularly convenient
since the radiation multipoles are then coordinate-independent. This allows the
equations of radiative transfer in a general spacetime to be recast
as multipole equations, which is essential for the transfer problem in
optically thin media. We applied the formalism to give a rigorous discussion of
the generation and propagation of CMB polarization in cosmological models.
We gave new results for the non-linear transformation properties of the
polarization multipoles under changes of reference frame, the exact source term
for Thomson scattering, and the multipole
propagation equations in linearised form in an almost-FRW model.
It was not necessary to split the
perturbations into scalar, vector and tensor modes to obtain these results,
and so they provide a solid foundation on which to build a complete,
second-order analysis of CMB polarization.
By expanding the linearised multipole equations in scalar
and tensor harmonics we derived the mode-expanded multipole equations
in almost-FRW models with general geometries, and their integral solutions.
These results confirm those obtained earlier by Hu~et al.~\cite{hu98}
with the total angular momentum method.

The linearised results of this paper have been included in a publically
available Fortran 90 code~\cite{LCL99}, based on CMBFAST~\cite{seljak96},
for calculating the CMB temperature anisotropy
and polarization in general FRW models, including those with closed
geometries, within the 1+3 covariant and gauge-invariant approach.

In a subsequent paper we will complete the development of a multipole
transfer formalism in general spacetimes by providing the exact
polarization multipole equations for an arbitrary geometry. This should
simplify the modelling of a number of important astrophysical situations
where relativistic flows or strong gravity effects are important,
including non-linear effects in the CMB.

\section*{Acknowledgements}

The author acknowledges a Research Fellowship from Queens' College, Cambridge.

\appendix

\section*{Scalar and Tensor Harmonics}
\label{sec:app}

In the text we make use of a specific PSTF representation of the scalar
and tensor harmonics. This representation was described in detail in
Ref.~\cite{chall99b}, so will only be summarised here. The starting point
is to ``coordinatise'' the solution in terms of a projected vector field
$e^a$, and a scalar field $\chi$. The restriction of these fields to the past
lightcone through some point $R$ (which can conveniently be taken to be
our point of observation) are the propagation direction of a free-streaming
photon which will subsequently pass through $R$, and the conformal look-back
time along the photon path respectively. The projected fields $e^a$ and $\chi$
are generated from their restriction to the past lightcone by Fermi
transporting along the integral curves of $u^a$:
\begin{equation}
\dot{e}^{\langle a \rangle} = 0, \qquad \dot{\chi}=0.
\label{eq:ap1}
\end{equation}
It is straightforward to show that the field $e^a$ satisfies
$e^a \uD_a e^b = 0$ at zero-order in an almost-FRW universe.

\subsection{Scalar harmonics}

In the PSTF representation the regular, normalisable scalar harmonics can be
written as
\begin{equation}
Q^{(k)} = \Phi_L^\nu(x) \clq_{A_L}^{(LM)} e^{A_L}, \qquad L \geq 0,
\label{eq:ap2}
\end{equation}
where the $\clq_{A_L}^{(LM)}$ with $M=-L \dots L$ are rank-$L$ PSTF tensor
fields satisfying the zero-order equations
\begin{equation}
e^b \uD_b \clq_{A_L}^{(LM)} = 0, \qquad
\dot{\clq}_{\langle A_L \rangle}^{(LM)}=0,
\label{eq:ap3}
\end{equation}
which determine the fields from their initial values at $R$. The
$\Phi_L^\nu(x)$ are ultra-spherical Bessel functions
(see, e.g. Ref.~\cite{abbott86}) with $\nu^2 = (k^2 + K)/|K|$ and
$x=\surd |K| \chi$. In closed models $\nu$ is restricted to integer values
with $\nu > L$. For convenience we label the harmonics with a lumped
superscript $(k)$ which represents $\nu$, $L$ and $M$.

The scalar harmonics are normalised so that
\begin{equation}
\int_0^\infty \ud \Omega_{e^a |_R} \ud x \, \sinh^2 \! x Q^{(k)}Q^{(k')}
= \frac{\pi}{2} \delta_{LL'}\Delta_L \clq_{A_L}^{(LM)}\clq^{(LM')A_L}
\nu^{-2} \delta(\nu-\nu'),
\label{eq:ap4}
\end{equation}
where $\ud\Omega_{e^a |_R}$ denotes an integral over solid angles
at $R$. Eq.~(\ref{eq:ap4}) refers to an open universe. In closed models
the hyperbolic functions replace their trigonometric counterparts,
$\delta(\nu - \nu')$ replaces $\delta_{\nu \nu'}$, and the upper limit on the
normalisation integral becomes $\pi$.
It is very convenient
to choose the $\clq^{(LM)}_{A_L}$ at $R$ so that
\begin{equation}
\clq_{A_L}^{(LM)}\clq^{(LM')A_L} = \Delta_L{}^{-1} \delta_{MM'},
\label{eq:ap5}
\end{equation}
which implies
\begin{equation}
\sum_{M=-L}^L \clq_{A_L}^{(LM)} \clq^{(LM)B_L} = \Delta_L{}^{-1}
h_{\langle A_L \rangle}^{\langle B_L \rangle}.
\label{eq:ap6}
\end{equation}
It follows from Eq.~(\ref{eq:ap3}) that Eqs.~(\ref{eq:ap5}) and (\ref{eq:ap6})
hold at zero-order at all points. A scalar-valued, statistically homogeneous
random field, say $\Delta(x^a)$, can be constructed from a superposition of
the scalar harmonics:
\begin{equation}
\Delta(x^a) = \int_0^\infty |K|^{3/2} \nu^2 \ud \nu \, \sum_{L=0}^\infty
\sum_{M=-L}^L \Delta_{\nu LM} Q^{(k)}, 
\label{eq:ap7}
\end{equation}
in an open universe. In the closed case the integral is replaced by a
discrete sum over integer $\nu$, and the sum over $L$ is restricted to
$L < \nu$. In the text we denote the sum over scalar harmonic modes
in the symbolic form $\Delta(x^a) = \sum_k \Delta_k Q^{(k)}$.
The covariance structure
\begin{equation}
\langle \Delta_{k} \Delta_{k'} \rangle =
\Delta^2(\nu) \delta_{kk'}
\label{eq:ap8}
\end{equation}
is sufficient to ensure statistical homogeneity and isotropy of the ensemble.
The symbolic $\delta_{kk'}$ represents $\delta_{LL'}\delta_{MM'}|K|^{-3/2}
\nu^{-2}\delta(\nu-\nu')$ in open models. For closed models $\delta(\nu-\nu')$
should be replaced by $\delta_{\nu\nu'}$.

From the $Q^{(k)}$ we can derive rank-$l$ PSTF tensors $Q^{(k)}_{A_l}$
as in Eq.~(\ref{eq:52}). For the specific representation given here we find
that at the point $R$,
\begin{equation}
Q^{(k)}_{A_{l}}|_R = \frac{1}{4\pi} \Delta_l \alpha_l \clq_{A_L}^{(LM)}
\delta_{lL},
\label{eq:ap9}
\end{equation}
so that the only modes to contribute to the $l$-th multipole of the
radiation anisotropy and polarization at $R$ have $l$ units of orbital angular
momentum. In a closed universe, the $Q^{(k)}_{A_l}$ vanish globally for
$l \geq \nu$.

\subsection{Tensor harmonics}

\subsubsection{Electric parity}

The regular, normalisable tensor harmonics with electric parity are given by
\begin{eqnarray}
Q^{(k)}_{ab} &=& T_1(x) \left(e_a e_b \clq_{C_L}^{(LM)} e^{C_L} +
\frac{1}{2} \clh_{ab} \clq_{C_L}^{(LM)} e^{C_L} \right) \nonumber \\
&& + T_2(x) e_{(a} \clh_{b)}^{c_L}
\clq_{C_L}^{(LM)} e^{C_{L-1}} + T_3(x) [\clq_{ab C_{L-2}}^{(LM)}
e^{C_{L-2}}]^{\TT}, \qquad L \geq 2.
\label{eq:ap10}
\end{eqnarray}
Here, the (screen) projection tensor $\clh_{ab}=h_{ab} + e^a e^b$, and the
subscript TT denotes the transverse (to $e^a$), trace-free part.
In an open universe, $T_1(x)$ is given by
\begin{equation}
T_1(x) = \frac{1}{\nu\sqrt{2(\nu^2+1)}}\left[\frac{(L+2)!}{(L-2)!}\right]^{1/2}
\frac{\Phi_L^\nu(x)}{\sinh^2\!x},
\label{eq:ap11}
\end{equation}
and $T_2(x)$ and $T_3(x)$ are determined by
\begin{eqnarray}
T_2(x) &=& \frac{-2}{(L+1)\sinh^2\!x} \frac{\ud}{\ud x}
[\sinh^3\!x T_1(x)],\label{eq:ap12} \\
T_3(x) &=& \frac{-L}{(L+2)} T_1(x) -
\frac{1}{(L+2) \sinh^2\!x} \frac{\ud}{\ud x}[\sinh^3\!x T_2(x)].
\label{eq:ap13}
\end{eqnarray}
The electric parity tensor harmonics are normalised so that
\begin{equation}
\int_0^\infty \ud \Omega_{e^a|_R} \ud x\sinh^2\!x \,
Q^{(k)}_{ab} Q^{(k')ab}
= \frac{\pi}{2}\delta_{LL'} \Delta_L \clq_{A_L}^{(LM)}
\clq^{(LM')A_L} \nu^{-2} \delta(\nu - \nu').
\label{eq:ap14}
\end{equation}
For closed universes, the hyperbolic functions should be replaced by their
trigonometric counterparts, $\nu^2+n$ should be replaced by $\nu^2-n$ with
$n$ an integer, and $\delta(\nu - \nu')$ should be replaced by
$\delta_{\nu \nu'}$. For closed models, the regular, normalisable modes have
$\nu$ an integer $\geq 3$, restricted to $\nu > L$.

We can form rank-$l$ ($l\geq 2$) PSTF tensors $Q^{(k)}_{A_l}$ from the electric
parity tensor harmonics as in Eq.~(\ref{eq:73}). At the point $R$ we find that
\begin{equation}
Q^{(k)}_{A_{l}}|_R = \frac{1}{4\pi}\frac{\sqrt{\nu^2+3}}{\nu}
M_l{}^{-1}\Delta_l \beta_l \clq_{A_L}^{(LM)} \delta_{lL},
\label{eq:ap15}
\end{equation}
so that, as with scalar modes, the only contribution to the $l$-th
multipole of the anisotropy and polarization at $R$ comes from those modes
with $l$ units of orbital angular momentum. In closed models the
$Q^{(k)}_{A_l}$ vanish for $l \geq \nu$.

\subsubsection{Magnetic parity}

For the magnetic parity tensor harmonics, which we denote with an overbar,
the regular normalisable solutions are
\begin{equation}
\bar{Q}^{(k)}_{ab} = \bar{T}_1(x) [e_{c_L} {\epsilon^{c_L
c_{L-1}}}_{(a}\clq_{b)C_{L-1}}^{(LM)} e^{C_{L-2}}]^{\TT}
+ \bar{T}_2(x) e_d e_{(a} {\epsilon_{b)}}^{d c_L} \clq_{C_L}^{(LM)}
e^{C_{L-1}},
\label{eq:ap16}
\end{equation}
with $L \geq 2$. In an open universe, $\bar{T}_2(x)$ is given by
\begin{equation}
\bar{T}_2(x)=\frac{1}{(L+1)}\left[\frac{2}{(\nu^2+1)}\right]^{1/2}
\left[\frac{(L+2)!}{(L-2)!}\right]^{1/2} \frac{\Phi_L^\nu(x)}{\sinh\! x},
\label{eq:ap17}
\end{equation}
and $\bar{T}_1(x)$ is determined by
\begin{equation}
\bar{T}_1(x) = \frac{-1}{(L+2)\sinh^2\!x} \frac{\ud}{\ud x}
[\sinh^3\!x \bar{T}_2(x)].
\label{eq:ap18}
\end{equation}
In a closed universe we make the same replacements as for the electric parity
harmonics. The magnetic parity harmonics satisfy the same normalisation
condition, Eq.~(\ref{eq:ap14}), as the electric parity harmonics, to which they
are orthogonal. We can also form rank-$l$ PSTF tensors $\bar{Q}^{(k)}_{A_l}$
from the magnetic parity harmonics, as in Eq.~(\ref{eq:73}).
However, we now find that the
$\bar{Q}^{(k)}_{A_l}$ vanish at $R$, so that the magnetic
parity harmonics do not contribute to the anisotropy and polarization there.
(However, magnetic parity modes do contribute at points not on the integral
curve of $u^a$ that passes through $R$.) In closed models the
$\bar{Q}^{(k)}_{A_l}$ vanish globally for $l \geq \nu$.

The electric and magnetic parity tensor harmonics are related through the
curl operation (which is parity reversing). For the conventions adopted here,
we find
\begin{eqnarray}
\curl Q^{(k)}_{ab} &=& {\frac{k}{S}} \sqrt{1+{\frac{3K}{k^2}}}
\bar{Q}^{(k)}_{ab},
\label{eq:ap19}\\
\curl\bar{Q}^{(k)}_{ab}&=&{\frac{k}{S}}\sqrt{1+{\frac{3K}{k^2}}}
Q^{(k)}_{ab}. \label{eq:ap20}
\end{eqnarray}

\subsubsection{Statistically homogeneous tensor-valued random fields}

We can construct a statistically homogeneous and isotropic tensor-valued
random field by superposing the electric and magnetic parity tensor harmonics.
For example, for the shear tensor $\sigma_{ab}$, we write
\begin{equation}
\sigma_{ab}=\int_0^\infty |K|^{3/2} \nu^2 \ud \nu\, \sum_{L=2}^{\infty} 
\sum_{M=-L}^{L} \frac{k}{S}
(\sigma_{\nu LM} Q^{(k)}_{ab} + \bar{\sigma}_{\nu LM} \bar{Q}^{(k)}_{ab}),
\label{eq:ap21}
\end{equation}
in an open universe. (Equation~[\ref{eq:ap21}] can easily be generalised to
include supercurvature modes, if these are present in the initial conditions.)
In a closed universe the integral over $\nu$ is replaced by a sum over
integer $\nu\geq 3$, and the sum over $L>2$ is further restricted to $L< \nu$.
In the text we use the symbolic summation $\sum_k$ to represent
the sum over modes on the right-hand side of Eq.~(\ref{eq:ap21}). The
requirements of statistical homogeneity and isotropy restrict the
covariance structure of the $\sigma_k$ and $\bar{\sigma}_k$ to the form
\begin{eqnarray}
\langle \sigma_{k} \sigma_{k'} \rangle &=& \sigma^2(\nu) \delta_{kk'}
\nonumber \\
\langle \bar{\sigma}_{k} \bar{\sigma}_{k'} \rangle &=& \sigma^2(\nu)
\delta_{kk'} \nonumber \\
\langle \sigma_{k} \bar{\sigma}_{k'} \rangle &=& 0, \label{eq:ap22}
\end{eqnarray}
where the symbolic $\delta_{kk'}$ is the same as for scalar modes.
The shorthand $\sigma_k$ is used to represent $\sigma_{\nu LM}$, and
similarly for the barred variables, in Eq.~(\ref{eq:ap22}) and also in the
text.

\end{document}